\documentclass[aps,reprint, prl, superscriptaddress, showpacs,nobibnotes]{revtex4-2}
\usepackage{graphicx}
\usepackage{hyperref}
\usepackage{bm,amsmath}
\usepackage{dcolumn}
\usepackage{natbib,xcolor}
\usepackage{gensymb}
\usepackage[version=4]{mhchem}

\newcommand{\ve}[1]{{\bf #1}}

\begin{document}

\title{Nonvolatile Nematic Order Manipulated by Strain and Magnetic Field in a Layered Antiferromagnet}
\author{Zili Feng}
\affiliation{Division of Physics, Mathematics and Astronomy, California Institute of Technology, Pasadena, CA 91125, USA}
\affiliation{Institute of Quantum Information and Matter, California Institute of Technology, Pasadena, CA 91125, USA}
\author{Weihang Lu}
\affiliation{Department of Physics and Astronomy, University of California Irvine,
Irvine, CA 92697, USA}
\author{Tao Lu}
\affiliation{Division of Physics, Mathematics and Astronomy, California Institute of Technology, Pasadena, CA 91125, USA}
\affiliation{Institute of Quantum Information and Matter, California Institute of Technology, Pasadena, CA 91125, USA}
\author{Fangyan Liu}
\affiliation{Department of Physics and Astronomy, University of College London, London WC1E 6BT, United Kingdom}
\affiliation{Division of Physics, Mathematics and Astronomy, California Institute of Technology, Pasadena, CA 91125, USA}
\affiliation{Institute of Quantum Information and Matter, California Institute of Technology, Pasadena, CA 91125, USA}
\author{Joseph R. Sheeran}
\affiliation{Division of Physics, Mathematics and Astronomy, California Institute of Technology, Pasadena, CA 91125, USA}
\author{Mengxing Ye}
\affiliation{Department of Physics and Astronomy, University of Utah, Salt Lake City, UT 84112, USA}
\author{Jing Xia}
\affiliation{Department of Physics and Astronomy, University of California Irvine,
Irvine, CA 92697, USA}
\author{Takashi Kurumaji}
\email{kurumaji@caltech.edu}
\affiliation{Division of Physics, Mathematics and Astronomy, California Institute of Technology, Pasadena, CA 91125, USA}
\affiliation{Institute of Quantum Information and Matter, California Institute of Technology, Pasadena, CA 91125, USA}
\author{Linda Ye}
 \email{lindaye@caltech.edu}
\affiliation{Division of Physics, Mathematics and Astronomy, California Institute of Technology, Pasadena, CA 91125, USA}
\affiliation{Institute of Quantum Information and Matter, California Institute of Technology, Pasadena, CA 91125, USA}

\date{\today}

\begin{abstract}
The operation mechanism of nematic liquid crystals lies in the control of their optical properties by the orientation of underlying nematic directors.
In analogy, electronic nematicity refers to a state whose electronic properties spontaneously break rotation symmetries of the host crystalline lattice, leading to anisotropic electronic properties.
In this work, we demonstrate that the layered antiferromagnet \ce{CoTa3S6} exhibits a switchable nematic order, evidenced by the emergence of both resistivity anisotropy and optical birefringence. This nematic state sets in at a temperature $T^*$ distinct from that of the antiferromagnetic transitions in the system, indicating a separate symmetry-breaking mechanism. The nematic order can be manipulated either by an in-plane rotation symmetry-breaking strain or in-plane magnetic field, with the latter exhibiting a pronounced non-volatile memory effect. Remarkably, we find that the broken three-fold rotation symmetry in electronic transport is restored with a moderate out-of-plane field. We hypothesize that the nematicity is of electronic origin and emerges from instabilities associated with van Hove singularities. The resulting phase diagram points to an intertwined interplay between the electronic nematicity and the proposed underlying collinear and non-coplanar spin orders.
Our findings establish \ce{CoTa3S6} as a versatile antiferromagnetic platform with highly tunable functionalities arising from the breaking of rotational, time-reversal, and inversion symmetries.

\end{abstract}

\maketitle

\begin{figure*}[t]
    \centering
    \includegraphics[width=0.85\textwidth]{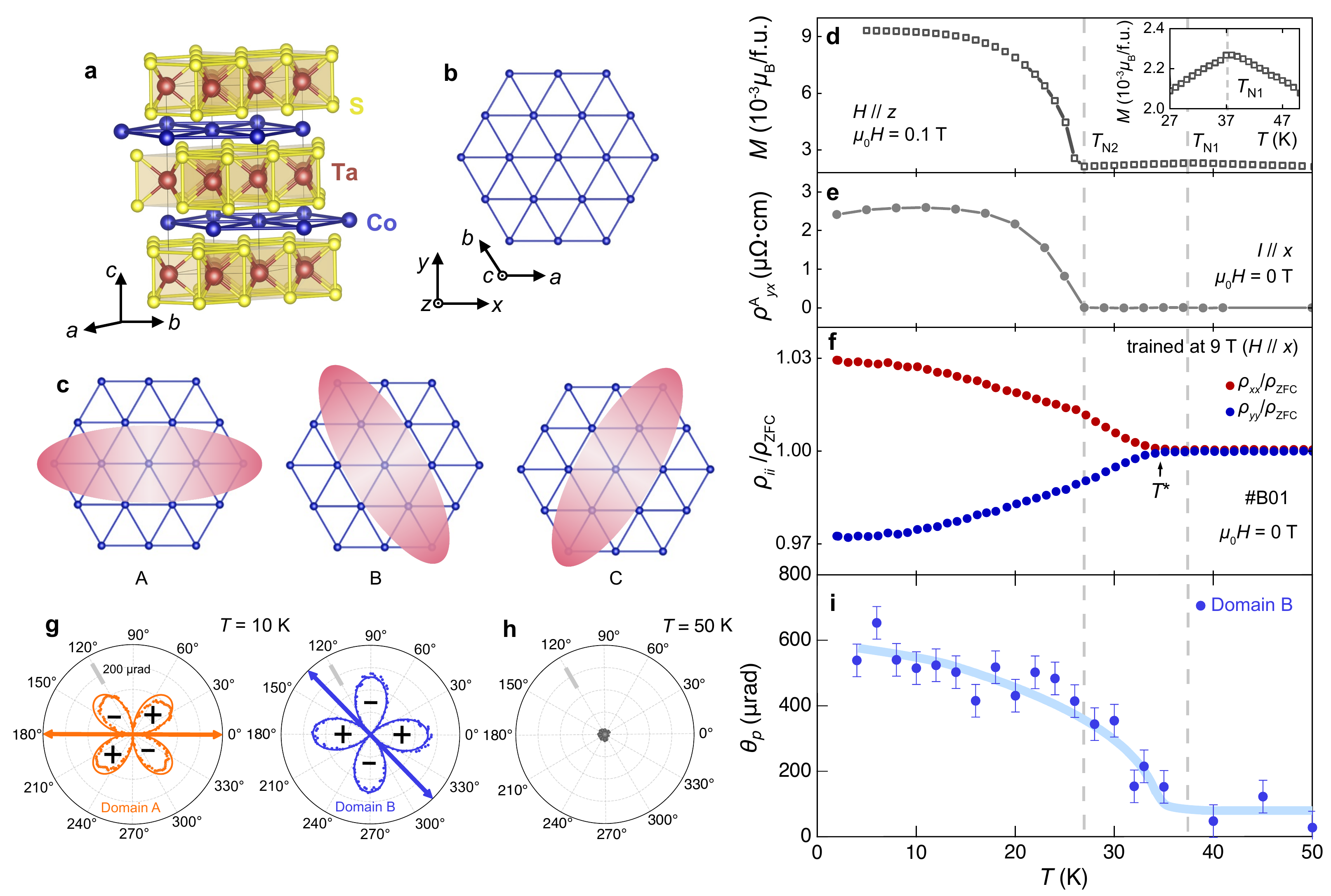}   
    \caption{\textbf{Emergence of resistivity anisotropy and optical birefringence in \ce{CoTa3S6}.}
    (a) Crystal structure of \ce{CoTa3S6}. Co atoms (blue spheres) are sandwiched by \ce{TaS2} layers composed of \ce{TaS6} triangular prisms. Ta (S) atoms are shown in red (yellow).
    (b) The Co-triangular lattice in the $ab$-plane;
    throughout the work we adopt a Cartesian coordinate system ($x,y,z$) and the definition for the axes with respect to the principal crystallographic axes is also shown.
    (c) Schematic of three-state nematicity on the Co triangular lattice; each of the inequivalent nematic domains is labeled as A, B and C, respectively.
    The long axes of the red shaded ellipses correspond to the direction along which $\Delta \rho > 0$.
    (d-f) Temperature $(T)$ dependence of magnetization (d), anomalous Hall resistivity $\rho^A_{yx}$ (e), and normalized resistivity $\tilde{\rho}_{ii}=\rho_{ii}/\rho_{\text{ZFC}}$ for $i=x,y$ (f).
    $\tilde{\rho}$ is defined as the ratio between field-trained and zero field cooled states (see text and Methods).
    Inset of (d) shows a magnified view of $M(T)$ near $T_{N1}$.
    $\rho^A_{yx}$ shown in (e) is the spontaneous Hall resistivity at each $T$ with scanning $H\parallel z$. $T^*$ in (f) indicates the onset temperature of resistivity anisotropy.
    (g–h) Polar plots of optical polarization rotation $\theta_T$ as a function of incident light polarization for the region of domain A and B at 10 K (g) and for domain B at 50 K (h).
    Plus/minus are the signs of the polarization rotation by birefringence.
    Solid curves are fits to $\theta_T = \theta_P \sin[2(\alpha - \alpha_0)]$ with $\theta_P,\alpha$,$\alpha_0$ all fitting parameters; the two-way arrows indicate the slow axis.
    (i) Fitted birefringence amplitude $\theta_P$ from measurements at various temperatures in domain B. The blue solid line is a guideline to the eye.}
    \label{fig:figure1}
\end{figure*}

Antiferromagnets, characterized by magnetic orders with zero or nearly zero net moments, have emerged as promising candidates for next-generation data storage devices due to minimal stray fields, robustness against magnetic field fluctuations, and fast switching dynamics \cite{baltz2018antiferromagnetic,jungwirth2016antiferromagnetic,vsmejkal2018topological}. 
However, the same set of attributes also pose significant challenges in manipulating antiferromagnetic domains, in stark contrast to ferromagnets, where magnetic fields—serving as the conjugate variable to the order parameter—enable straightforward writing of magnetic bits.  
The realization of functional memory devices based on antiferromagnets thus hinges on identifying robust, non-volatile orders that can be manipulated by external stimuli and reliably read by external probes.
Recent advances include electrical readout of antiferromagnetic Néel vectors \cite{Marti2014-mx,Wadley2016-fb} and their current-switching in non-centrosymmetric antiferromagnetic systems \cite{Zelezny2014-zx,Wadley2016-fb,Grzybowski2017-ou,Nair2020-ts}. From a band structure perspective, antiferromagnets also allow a versatile interplay between various symmetries with topology and geometry of the underlying electronic states \cite{vsmejkal2018topological,Gao2023-au,PhysRevLett.133.106701}.

As a distinct class of quantum phases, electronic nematicity refers to states where the electronic responses spontaneously break rotational symmetries of the underlying crystalline lattice, and can be viewed as a quantum analog of the orientational orders in liquid crystals \cite{Kivelson1998-ki,fradkin2010nematic}.
Initially identified in two-dimensional electron gas at the quantum Hall regime \cite{lilly1999evidence}, nematicity has been reported in various strongly correlated oxides and pnictides \cite{Borzi2007-li,Hinkov2008-dh,Chu2010-cl,fradkin2010nematic,bohmer2022nematicity,Fernandes2022-vl}, and most recently in moiré superlattices \cite{Cao2021-sv,Rubio-Verdu2022-tm} and kagome superconductors \cite{Xu2022-hu,Li2023-ag,Yang2024-kf}. A nematic state selects a special direction in space, often indicated by a headless vector termed nematic director, while the particular rotation symmetry that's broken determines the number of domains which would in principle serve as memory bits.
Examples include Ising nematicity with two distinct states \cite{Kivelson1998-ki,bohmer2022nematicity} and Potts nematicity with three \cite{Fernandes2020-us,little2020three,hwangbo2024strain}. Analogous to liquid crystals, the anisotropy induced by the nematic order, and its response with respect to external parameters are key to its functionalities. 

 Here we present an unexpected discovery of a unique three-state nematic state that coexists and interacts with underlying antiferromagnetic orders in a layered compound \ce{CoTa3S6}. The electronic anisotropy of the system is found to be closely intertwined with the antiferromagnetic orders. 
\ce{CoTa3S6} belongs to a family of magnetic element-intercalated transition metal dichalcogenides ($T$\ce{X2}, $T$=Nb, Ta, $X$=S, Se) that were first studied in the 1970s driven by the rich magnetic states that can be realized depending on the intercalant species and concentration \cite{parkin19803A,parkin19803B,miyadai1983magnetic,morosan2007sharp,wu2022highly,xie2022structure}. \ce{CoTa3S6} can be viewed as an ordered structure for fractional (1/3) intercalation of Co into the van der Waals gap between \ce{TaS2} layers forming a $\sqrt{3}\times\sqrt{3}$ superlattice. The crystal structure is shown in Fig.~\ref{fig:figure1}a, while in reality a small off-stoichiometry may be allowed \cite{park2024composition}. The cobalt atoms between the \ce{TaS2} layers form triangular lattices (Fig.~\ref{fig:figure1}b) and the overall crystal structure lacks inversion symmetry and is chiral, belonging to the space group $P6_322$ \cite{VANDENBERG1968143}.

A recent surge of attention on \ce{CoTa3S6} and isostructural \ce{CoNb3S6} was stimulated by reports of large anomalous Hall effects in the $xy$-plane arising in both compounds within their antiferromagnetic states \cite{Ghimire2018-bn,park2022field,PhysRevB.103.184408,PhysRevResearch.2.023051,tanaka2022large}.
 For \ce{CoTa3S6} in particular, two successive magnetic transitions are reported at $T_{N1} \approx$ 37 K and $T_{N2} \approx$ 26 K, while the anomalous Hall effect is exclusively observed below $T_{N2}$ \cite{parkin19803A,parkin19803B,park2022field}.
 Neutron scattering studies have revealed magnetic modulation vectors $\{1/2,0,0\}$ below both $T_{N1}$ and $T_{N2}$, and a single-$q$ collinear antiferromagnetic state is proposed between $T_{N1}$ and $T_{N2}$ while a four-sublattice non-coplanar spin structure is proposed below $T_{N2}$ \cite{park2023tetrahedral,takagi2023spontaneous,Park2024-nl}.
 The latter as a triple-$q$ state is supported by the observation of higher harmonics $\{1/2,1/2,0\}$ and symmetry analysis of magnetic point group to account for the observed anomalous Hall effect \cite{park2023tetrahedral,takagi2023spontaneous,yanagi2023generation,park2024dft+,heinonen2022magnetic}.
 That anomalous Hall effect is allowed here suggests that the system is a promising candidate for multi-functional, antiferromagnetic spintronic applications.

 \begin{figure*}[t]
    \centering
   \includegraphics[width=0.8\textwidth]{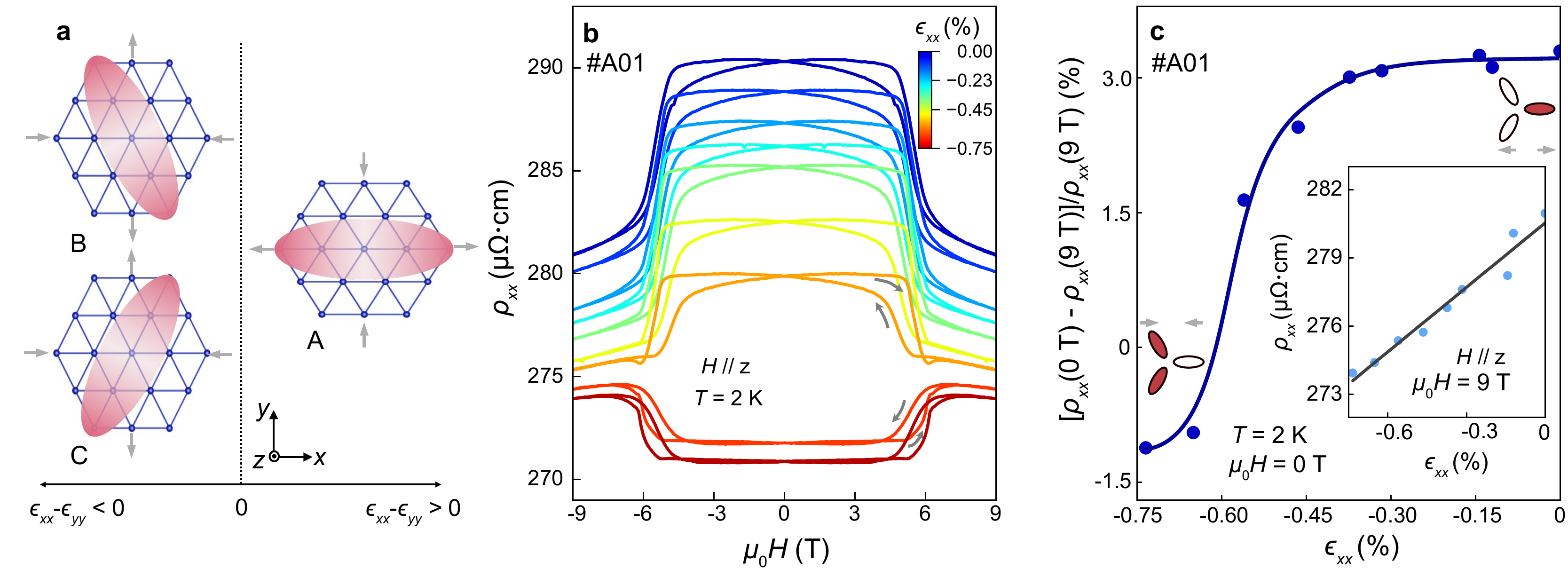}
    \caption{\textbf{Resistivity anisotropy and nematic domains manipulated by in-plane strain.}
    (a) Schematic strain effect on the Co triangular lattice and expected nematic domain selection by in-plane strain.
    (b) Out-of-plane magnetic field ($H\parallel z$) dependence of $\rho_{xx}$ measured at 2 K under various magnitudes of strain along the $x$ axis ($\epsilon_{xx}$).
    (c) Strain dependence of the magnetoresistance MR = ($\rho_{xx}$(0 T) - $\rho_{xx}$(9 T))/$\rho_{xx}$(9 T) (blue circle). Dark blue curve is a guideline to the eye.  The expected activated nematic domains at the compressive and tensile ends of the strain range are also shown.
    Inset is the strain dependence of $\rho_{xx}$(9 T) fitted with a linear function of $\epsilon_{xx}$.}
    \label{fig:figure2}
\end{figure*}

\noindent\textbf{Identification of nematic order in \ce{CoTa3S6}} We highlight our central observation of anisotropic electronic behaviors in \ce{CoTa3S6}--which allows identification of a three-state nematic state (Figs.~\ref{fig:figure1}c) --in close comparison with the magnetic transitions in the system in Figs.~\ref{fig:figure1}d-i.
Magnetization ($M$) along $z$ was measured with $H\parallel z$ at $\mu_0H=0.1$ T (Fig.~\ref{fig:figure1}d); a weak kink can be observed at $T_{N1}=37$ K (see Fig. ~\ref{fig:figure1}d inset) while an onset of a net moment can be observed below $T_{N2}=27$ K. Importantly, we note that the net moment is two orders of magnitude smaller than that expected for a ferromagnetic alignment of Co$^{2+}$ moments ($\sim 3.8 $ $\mu_ B$). The onset of the net moment below $T_{N2}$ is also accompanied by the appearance of a spontaneous anomalous Hall effect (Fig. ~\ref{fig:figure1}e), comparable with that reported in previous studies \cite{park2023tetrahedral,takagi2023spontaneous} (see Supplementary Information (SI) Sec. I for characterization of our crystals from the anomalous Hall effect).

The electronic anisotropy is sensitively captured through in-plane transport as exemplified by the temperature dependence of the normalized resistivities $\tilde{\rho}_{xx}$ and $\tilde{\rho}_{yy}$ in Fig.~\ref{fig:figure1}f.
The $x$ axis is along the bonds of the Co triangular lattice as shown in Fig. ~\ref{fig:figure1}b.
Here, $\tilde{\rho}_{ii}(T)$ is defined by $\tilde{\rho}_{ii}=\rho_{ii}(T,\text{FC})/\rho_{ii}(T,\text{ZFC})$; $\rho_{ii}(T,\text{FC})$ is measured at zero field after the field-trained process at 9 T of $H\parallel x$ (we will also get back to the training procedure below), while $\rho_{ii}(T,\text{ZFC})$ is measured after zero-field cooling. 
$\rho_{xx}$ and $\rho_{yy}$ are obtained through simultaneous measurements using a modified Montgometry method (see Methods and SI Sec. IV).
Note that for a hexagonal crystalline lattice, no resistivity anisotropy is expected, i.e., $\tilde{\rho}_{xx}=\tilde{\rho}_{yy}=1$.

We have found that the in-plane resistivity anisotropy turns on below $T_{N1}$ at $T^*\simeq34$ K, which persists to low temperature.
Recalling that the measurements are performed at zero field, it indicates that the six-fold rotation symmetry of the crystal is spontaneously broken below the characteristic temperature scale $T^*$.
Furthermore, the antisymmetric evolution of $\tilde{\rho}_{xx}$ and $\tilde{\rho}_{yy}$ with temperature implies that the transport tensor indeed develops a nematic-like anisotropy in the $xy$ plane with $\rho_{xx}=\bar{\rho}+\Delta\rho,\rho_{yy}=\bar{\rho}-\Delta\rho$, where $\bar{\rho}$ is the average in-plane resistivity and $\Delta\rho$ reflects the difference in $\rho_{xx}$ and $\rho_{yy}$ relative to the corresponding isotropic state $(\rho_{xx}=\rho_{yy}=\bar{\rho})$. We take the convention that the long axes of the nematic directors (Fig. \ref{fig:figure1}c) has $\Delta\rho>0$.
We have verified that an isotropic, multi-domain state is realized in $\rho(\text{ZFC})$ (see SI Sec. IV), and expect $\tilde{\rho}_{xx}-\tilde{\rho}_{yy}=2\Delta\rho/\bar{\rho}$ to play the role of a nematic order parameter.
We note that a very weak kink in resistivity anisotropy may be identified near the onset of net moment/anomalous Hall effect at $T_{N2}$.

The emergence of electronic anisotropy in \ce{CoTa3S6} is further evidenced by the observation of optical birefringence at low temperature (see Methods and SI Sec. II).
In Fig. \ref{fig:figure1}g we show the polar plot of the birefringence response at 10 K of two domains we have identified. Their optical principal axes are close to that expected for domains A and B illustrated in Fig. \ref{fig:figure1}c, respectively.
We find that the slow axis tends to align with the $a$ or equivalent crystallographic axes. 
These rotation symmetry-breaking patterns are in contrast to an isotropic state at 50 K (Fig. \ref{fig:figure1}h). The $T$-evolution of birefringence of domain B is summarized in Fig. \ref{fig:figure1}i, which closely follows the $T$-evolution of resistivity anisotropy (Fig. \ref{fig:figure1}f); both highlighting a growing electronic anisotropy with lowering $T$.

That electronic anisotropy emerges at a temperature between $T_{N1}$ and $T_{N2}$ and persists in the triple-$q$ ground state is rather unexpected, as the superposition of three equivalent $q$ vectors is expected to retain the three-fold rotational symmetry \cite{park2023tetrahedral,takagi2023spontaneous}.
These all point to that the nematicity in \ce{CoTa3S6} likely is a distinct phase from the antiferromagnetic orders; we will return to this below.\\

\begin{figure*}[th]
    \centering    \includegraphics[width=0.85\linewidth]{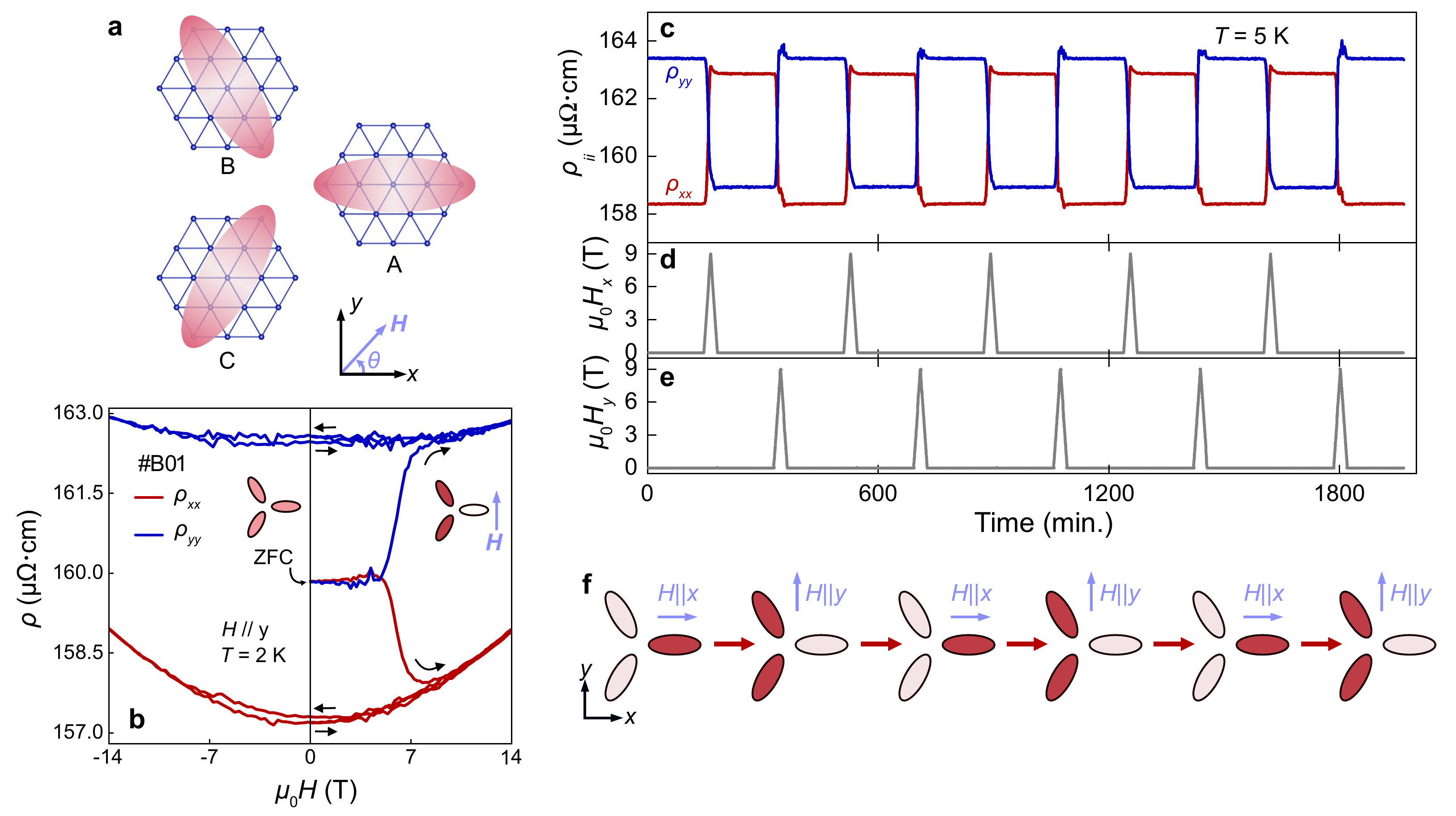}
    \caption{\textbf{Manipulation of resistivity anisotropy by in-plane magnetic fields.}
    (a) Schematic of the relationship between in-plane magnetic field direction and the nematic director.
    $\theta$ is the angle of the magnetic field with respect to the $x$ axis. 
    (b) $\rho_{xx}$ and $\rho_{yy}$ as a function of in-plane magnetic field along $y$-axis ($\theta =90^{\circ}$).
    The initial state is prepared by ZFC from $T = 50$ K and 2 K.
    The magnetic field is applied in the sequence of $\mu_0H=0$ T $\rightarrow +14$ T $\rightarrow -14$ T $\rightarrow +14$ T.
    (c-e) Time-dependence of $\rho_{xx}$ and $\rho_{yy}$ (c) with subsequent application of $H_x$ and $H_y$ training fields with field history as a function of time shown in (d) and (e).
    (f) Schematic of domain selection with in-plane magnetic field alternating between $H_x$ and $H_y$.
    Ellipses with darker shade indicate the favored domain(s) at respective field configurations. 
}
    \label{fig:figure3}
\end{figure*}

\noindent\textbf{Controlling nematic domains with in-plane strain} With a finite coupling between an anisotropic, nematic state and the lattice, one may expect using in-plane strain that breaks rotation symmetry in the same manner of the nematic order parameters to couple to the former and control the nematic domains \cite{Chu2010-cl,Chu2012-vj,little2020three,hwangbo2024strain}. 
In Fig. \ref{fig:figure2}a we illustrate a possible scenario of domain selection (the validity of the scenario is corroborated by experiments discussed below) as a function of in-plane antisymmetric strain $\frac {1}{2}\left(\epsilon_{x x}- \epsilon_{y y}\right)$: with $\frac {1}{2}\left(\epsilon_{x x}- \epsilon_{y y}\right)>0$ one out of the three Potts nematic domains is selected while for $\frac {1}{2}\left(\epsilon_{x x}- \epsilon_{y y}\right)<0$ two out of the three nematic domains are selected. In our experiment, we apply a uniaxial stress along $x$ which generates a finite component in $\frac {1}{2}\left(\epsilon_{x x}- \epsilon_{y y}\right)$. The longitudinal resistivity $\rho_{xx}$ at different bias strains $\epsilon_{xx}\equiv \Delta L_x/L_x$ are summarized in Fig. \ref{fig:figure2}b, and here the magnetic field is applied along the $z$-axis.

Qualitatively, as a function of $\epsilon_{xx}$, the magnetoresistance in Fig. \ref{fig:figure2}b appears to abruptly switch sign from negative on the most tensile side (blue curves) to positive on the most compressive side (red curves).
This can also be seen in the $\epsilon_{xx}$-dependence of both the zero field and high field $\rho_{xx}$ summarized in Fig. \ref{fig:figure2}c.
At 9 T, $\rho_{xx}$ appears to evolve linearly with $\epsilon_{xx}$ (Fig. \ref{fig:figure2}c inset), and similar linear elastoresistivity can arise from a combination of the strain gauge effect and modification of the band structure under strain \cite{amin1991elastoresistance,mutch2019evidence}.
To eliminate such contributions in the strain-evolution of $\rho_{xx}$(0 T), we compare the percentage magnetoresistance $\left[\rho_{x x}(0 \mathrm{~T})-\rho_{x x}(9 \mathrm{~T})\right] / \rho_{x x}(9 \mathrm{~T})$ against $\epsilon_{x x}$ in Fig.~\ref{fig:figure2}c. 
This quantity can be found to be highly nonlinear with $\epsilon_{xx}$, and it changes sign near $\epsilon_{xx}=-0.6\%$ and tends to saturate towards both the compressive and tensile ends of $\epsilon_{xx}$ (see SI Sec. V for similar results at 10 K). 

As demonstrated above in Fig. \ref{fig:figure1}f, $\rho_{xx}$ and $\rho_{yy}$ evolves in an antisymmetric manner within the nematic state; this allows us to use either $\rho_{xx}$ and $\rho_{yy}$ on their own as an indication of the nematic director. 
We therefore attribute the different $\rho_{xx}$ values at the compressive/tensile sides to a resistive detection of the strain-induced selection of nematic directors (see schematics in Fig. \ref{fig:figure2}c). The dark blue curve in Fig. \ref{fig:figure2}c is a guideline to the eye. We note that in the ideal case, in-plane strain tuning of resistivity anisotropy is expected to exhibit a linear interpolation between values at the two ends, as $\epsilon_{xx}-\epsilon_{yy}$ is a thermodynamic variable expected to be proportional to imbalance of the nematic domains \cite{Bartlett2021-hv}.\\ 

\noindent\textbf{In-plane magnetic field manipulation of nematic domains} Having demonstrated that in-plane strain can control the nematic domains, next we turn to the manipulation of nematic domains by in-plane magnetic fields \cite{Lilly1999-ix,Borzi2007-li,bohmer2022nematicity}. Similar with in-plane strains, the presence of an in-plane magnetic field breaks the six-fold rotation symmetry of the lattice and is expected to couple to and can act to select amongst three nematic domains, as schematically illustrated in Fig.~\ref{fig:figure3}a. 

Figure \ref{fig:figure3}b shows a typical in-plane field-evolution of resistivity anisotropy at $T$ = 2 K starting from a zero field cooled, multi-domain state.
Here the in-plane field is applied along $y$ $(\theta=90^{\circ})$.
The ZFC state is isotropic in $\rho_{xx}$ and $\rho_{yy}$, while as field increases, a finite resistivity anisotropy starts to emerge from 7 T with $\rho_{xx}$ and $\rho_{yy}$ again bifurcating in opposite, antisymmetric manners.
The weak parabolic magnetoresistance in both $\rho_{xx}$ and $\rho_{yy}$ can be attributed to the orbital Lorentz force. We note that as field reaches 14 T, the resistivity anisotropy saturates, and when the magnetic field is swept back and forth between $\pm 14$ T, the system maintains the resistivity anisotropy initialized at high magnetic fields.  This suggests that (1) in-plane field along $y$ acts to select B and C domains, both of which exhibits a resistivity anisotropy $\rho_{xx}-\rho_{yy}<$0 and (2) the memory of resistivity anisotropy and nematic domains is retained when the magnetic field is removed.

The non-volatile nature of the nematic domains allows one to utilize them as medium for a field-controlled nematic memory.
We therefore devise a writing procedure using pulse-like in-plane magnetic fields to train the domains and subsequently using resistivity anisotropy as a readout mechanism (Fig. \ref{fig:figure3}c-e).
At 5 K the switching between states with $\rho_{xx}-\rho_{yy}>0$ and $\rho_{xx}-\rho_{yy}<0$ are triggered by $x$ and $y$ magnetic fields, respectively, and the operation can be repeated multiple cycles without decay in amplitude.
Fig. \ref{fig:figure3}f illustrates the corresponding activated nematic domains in the in-plane field training sequence.
Such effect is found to persists up to 25 K (see SI Sec. VIII).
We note that similar non-volatile domain selection is expected for the application and removal of in-plane uniaxial stress, which is not accessed in the present set of experiments (see Methods). In SI Sec. IX, we further verify that the nematic director is pinned along the principal $a$ and equivalent axes of the crystal and can be rotated in a discrete manner by the rotation of the in-plane field. \\

\begin{figure*}
    \centering
    \includegraphics[width=0.9\linewidth]{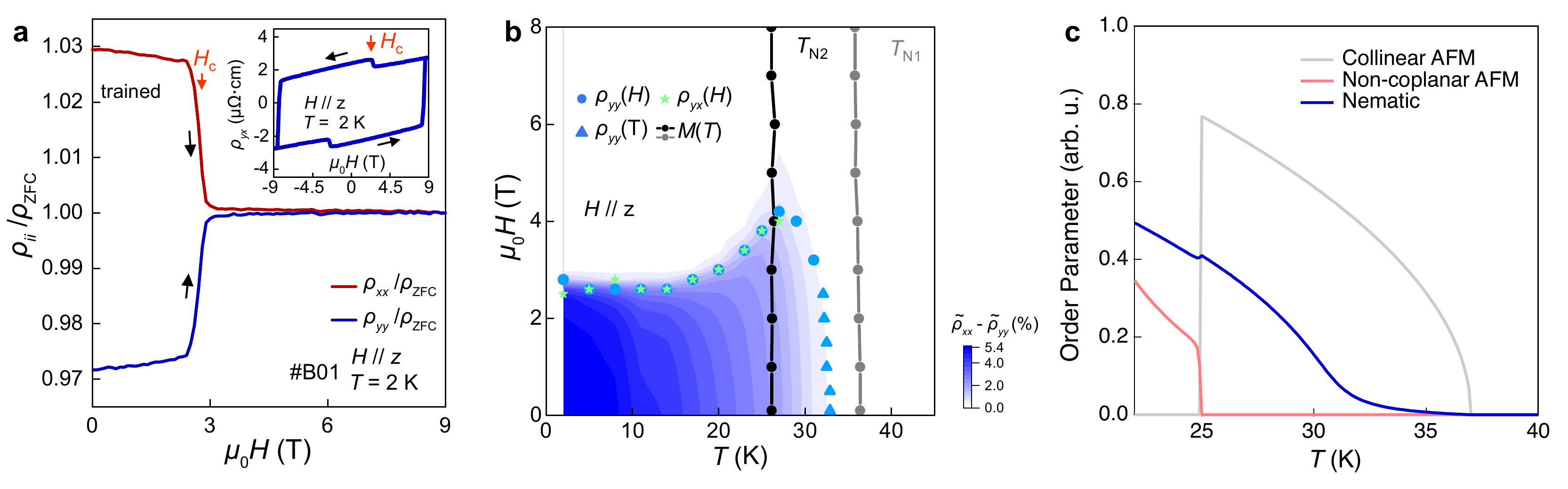}
    \caption{\textbf{Phase diagram of \ce{CoTa3S6} for magnetic field along the $z$-axis.}
    (a) Magnetic field dependence of normalized in-plane resistivities ($\tilde{\rho}_{ii}$) with $H\parallel z$ at $T = 2$ K.
    Prior to the measurement, the sample is field-cooled with in-plane magnetic field along the $x$ axis at 9 T (see text and Methods).
    The inset shows the magnetic field dependence of $\rho_{yx}$ at $T=2$ K.
    Red arrows indicate the critical field ($H_c$).
    (b) Color map of the resistivity anisotropy ($\tilde{\rho}_{xx}-\tilde{\rho}_{yy}$) in the $H\parallel z-T$ plane.
    The transition temperatures $T_{N1}$ (gray circles) and $T_{N2}$ (black circles) are extracted from the temperature dependence of magnetization. Blue circles (triangles) are defined by singularities in $\rho_{yy}(H)$ ($\rho_{yy}(T)$) scans (see SI Sec. IV and XI) and green stars are defined by kinks in $\rho_{yx}(H)$. (c) Temperature evolution of nematic, collinear and non-coplanar antiferromagnetic order parameters in a Landau free energy picture (see Methods and text).}
    \label{fig:figure4}
\end{figure*}

\noindent\textbf{Temperature-field phase diagram of \ce{CoTa3S6}} In order to gain additional insights on the origin of nematicity in relation to the underlying magnetic orders in \ce{CoTa3S6}, below we evaluate the effect of out-of-plane field to the observed transport anisotropy.
Figure ~\ref{fig:figure4}a shows $\tilde{\rho}_{ii}$ as a function of out-of-plane magnetic field ($H\parallel z$).
Here, $\tilde{\rho}_{ii}(H)$ is the normalized resistivity defined by $\tilde{\rho}_{ii}(H)=\rho_{ii}$($H$,FC)/$\rho_{ii}$($H$,ZFC); $\rho_{ii}$($H$,FC) is measured after field cooling with in-plane magnetic field from $T=50$ K to 2 K, while $\rho_{ii}$($H$,ZFC) is measured after zero-field cooling (see Methods). As we have demonstrated above, following the removal of in-plane training field, the system retains the memory of the nematic domains at low temperature. Starting from an anisotropic state thus prepared, as an out-of-plane field is gradually turned on, as shown in Fig.~\ref{fig:figure4}a, the resistivity anisotropy of the trained state suddenly collapses at a critical magnetic field $H_{c}\simeq 3$ T, which corresponds well with the sudden drop in $\rho_{yx}$ (inset of  Fig.~\ref{fig:figure4}a) and sudden jump in $M_z$ (see SI Sec. I) both observed in our own crystals and reported in literature \cite{park2022field,park2023tetrahedral,takagi2023spontaneous}. This sudden collapse of resistivity anisotropy suggests that $H_c$ is a characteristic field scale where the nematic order is undermined by a moderate out-of-plane field; in other words, at $H_c$ the system ascends to a rotation symmetric state. 

We note that above the critical field $H_c$, $\rho_{xy}$ is comparable with that at zero field (Fig. \ref{fig:figure4}a inset), while the resistivity anisotropy vanishes thoroughly. This suggests that the nematic order and the anomalous Hall-driving antiferromagnetic phase are likely of distinct origin. Notably, in experiments with canted magnetic fields (see SI Sec. VI), we find that the control of resistivity anisotropy is independent of the sign of the anomalous Hall effect, suggesting that the former is controlled by in-plane components while the latter by $z$-component of the magnetic field only. These different responses again indicate that the nematicity and Hall effect arise from distinct degrees of freedom in \ce{CoTa3S6}.

In Fig. \ref{fig:figure4}b we summarize the evolution of resistivity anisotropy in a color scale in the out-of-plane field and temperature plane, and solid marks indicate phase boundaries obtained from magnetization and transport ($\rho_{yy},\rho_{yx}$) measurements. At all temperatures below $T_{N1}$, the field scale where resistivity anisotropy vanishes agrees well with that defined by kinks in $\rho_{yx}$. 
Additionally, the shape of the phase diagram also implies that the nematic order below $T^*$ and $H_c$ is a distinct phase from the collinear antiferromagnetic order which onsets at $T_{N1}$ and the non-coplanar antiferromagnetic order which sets in at $T_{N2}$: not only does $T^*$ at zero field is distinct from both $T_{N1}$ and $T_{N2}$, but the $T^*$ nematic phase is much more fragile with respect to out-of-plane field; no discernible changes of $T_{N1}$ and $T_{N2}$ can be identified up to 9 T.\\

\noindent\textbf{Discussion} Thus far, our observations have established the presence of a three-state nematic state within the antiferromagnetic orders in \ce{CoTa3S6}. 
The state can be read out by resistivity anisotropy and optical birefringence and manipulated by in-plane strain and magnetic fields.
Additionally, we identified that the rotation symmetry-breaking can be `turned off' by an out-of-plane magnetic field, offering an additional means of tuning the anisotropic functionalities of the system. The fact that this suppression is also evident in both magnetization and anomalous Hall signals (see SI Fig. S2 and Fig. \ref{fig:figure4}a inset) suggests that, while the nematic order constitutes a distinct degree of freedom, it is coupled to the underlying magnetic phases.

Here we illustrate the interplay between the three orders in \ce{CoTa3S6} using a minimal Landau free energy picture. Here we consider $(\ve{L}_1,\ve{L}_2,\ve{L}_3)$ for the magnetic order parameter ($\ve{L}_i$ is a N\'{e}el vector for an antiferromagnetic configuration propagating along one of the $M$ points in the hexagonal Brillouin zone \cite{little2020three}) and $(n_1,n_2)=n(\cos2\theta,\sin2\theta)$ for the nematic order parameter \cite{little2020three,Fernandes2020-us}. The free energy associated with the magnetic orders is (see SI Sec. III for a more complete description)
\begin{align}
	F_{\rm mag}&= a \sum_{i=1}^{3} \ve{L}_i^2 + v_0 \left( \sum_{i=1}^{3} \ve{L}_i^2 \right)^2+ v_1 \sum_{i<j} \ve{L}_i^2 \ve{L}_j^2 \nonumber\\&+ v_2  \sum_{i<j} \left(\ve{L}_i \cdot \ve{L}_j\right)^2\label{FE1}
\end{align}
with the associated Landau parameters $a = \tilde{a} (T-T_{N1}), v_1 = \tilde{v}_1 (T-T_{N2}),  v_0>0, \tilde{a}>0, \tilde{v}_1 >0$. The competition between $v_1$ and $v_2$ terms drives the single-$q$ to triple-$q$ transition \cite{Ye2017-ax}. The free energy for the nematic state together with its coupling with the modulated magnetic states is
\begin{align}
	&F_{\rm n} + F_{\rm n-mag} = b n^2 + c n^3 \cos{6\theta} + w n^4 \nonumber\\&+ g n \left( \cos{2\theta} ( \ve{L}_1^2 + \ve{L}_2^2 - 2 \ve{L}_3^2) + \sin{2\theta}\sqrt{3} ( \ve{L}_1^2-\ve{L}_2^2)\right)\label{FE2}
\end{align}

With the above formalism (see Methods for the parameter values used), we are able to produce the schematic order parameter evolution in Fig. 4c.
The behaviors closely mirror the onset of the anomalous Hall effect (Fig. \ref{fig:figure1}e) and anisotropy (Fig. \ref{fig:figure1}f,i) of the system, as well as the weak cusp at $T_{N2}$ in the latter (Fig. \ref{fig:figure1}f and Fig. S13).
The free energy analysis also points to that above $T^*$, a tail in the nematic order parameter is induced by a finite coupling with the stripe antiferromagnet order at $T_{N1}$, which is captured in resistivity anisotropy (Fig. \ref{fig:figure1}f and Fig. S13).
The clear similarities imply that the nematic and the modulated magnetic order parameters are indeed distinct and intertwined orders, which contrasts with (vestigial) nematicity induced by stripe antiferromagnetic orders \cite{Fernandes2014-qg,little2020three,hwangbo2024strain}, or rotation-symmetry-breaking charge density waves \cite{Nie2014-sz,Nie2022-tx}.
The emergence of nematicity here as a separate, non-vestigial order is rather surprising; to our knowledge, \ce{CoTa3S6} represents a quite unique case of intertwined nematic and antiferromagnetic orders with distinct origins.

We hypothesize that the nematic order in \ce{CoTa3S6} is primarily driven by electronic rather than magnetic degrees of freedom. For instance, photoemission studies show hexagonal and triangular Fermi surfaces that nearly touches the $M$-points of the Brillouin zone \cite{park2023tetrahedral,Kar2025-jl}, indicating the presence of van Hove singularities near the Fermi level. Such van Hove singularities in hexagonal lattices have been widely discussed as a driving force of a wide variety of electronic orders, including nematicity through the Pomeranchuk instability \cite{PhysRevB.87.115135,PhysRevB.98.205151,PhysRevB.107.184504,Fernandes2020-us,Jiang2024-te,Nag2024-on}. This intertwined interplay of orders echoes paradigms seen in strongly correlated systems such as cuprates and iron pnictides \cite{RevModPhys.87.457,Fernandes2019-kt}, and more recently in kagome metals with itinerant electrons on frustrated lattices with electronic singularities \cite{Fernandes2019-kt,Teng2023-ay,PhysRevX.14.011043,Farhang2025-ob}. 
Elucidating the microscopic origin of the observed nematicity in \ce{CoTa3S6} represents a compelling direction for future research and may help establish design principles for electronic instabilities in metallic systems.

As a final remark, we note that structurally, \ce{CoTa3S6} lacks inversion symmetry through its chiral structure, and time-reversal symmetry is known to be broken at low temperature evidenced by the anomalous Hall effect. Our experimental results further reveal that the zero magnetic field ground state of \ce{CoTa3S6} breaks the underlying rotation symmetry of the lattice. The interplay between the various broken symmetries may lead to novel electronic and optoelectronic functionalities in \ce{CoTa3S6}, such as current-tunability of the nematic order and/or the anomalous Hall effect analogous to current switching of N\'{e}el vectors in \ce{Fe_{1/3}NbS2} \cite{Nair2020-ts}. Moreover, the observed non-volatile memory associated with the nematic order may enable control over nonlinear electrical transport \cite{Ideue2021-of} and spin current generation \cite{Yang2021-lu}, leveraging the simultaneous absence of inversion and time-reversal symmetries.

\subsection{Acknowledgement}

We thank Veronika Sunko and Patrick Lee for fruitful discussions. The experimental works carried out at Caltech are supported by Gordon and Betty Moore Foundation through Moore Materials Synthesis Fellowship to L.Y. (GBMF12765) and Institute of Quantum Information and Matter (IQIM), an NSF Physics Frontier Center (PHY-2317110). The optical measurements at University of California, Irvine are supported by Gordon and Betty Moore Foundation through Emergent Phenomena in Quantum Systems (EPiQS) Initiative Grant GBMF10276 to J.X.. M.Y. is supported by a start-up grant from the University of Utah. M.Y. and L.Y. acknowledge support from the Gordon and Betty Moore Foundation’s EPiQS Initiative (Grant GBMF11918), which enabled valuable discussions.
 Z.F. acknowledges support from IQIM Postdoctoral Fellowship at Caltech. Part of the work at Caltech is carried out at the X-Ray Crystallography Facility supported by the Beckman Institute and at the shared Physical Properties Measurement System facilities supported by NSF DMR-2117094.

\section{Methods}
\subsection{Crystal Growth and Characterization}
Single crystals of \ce{CoTa3S6} were grown by a two-step process \cite{parkin19803A,parkin19803B,park2022field,takagi2023spontaneous}.
Polycrystalline CoTa$_{3}$S$_{6}$ was first synthesized by solid state reaction from stoichiometric mixture of Co (99.99\%, 
Strem Chemicals), Ta (99.98\%, Strem Chemicals) and S (99.9995\%,Alfa Aesar) powders.
The mixture was loaded in an alumina crucible, which was then sealed in an evacuated quartz tube.
The tube was slowly heated up to 900\degree C and kept for one day.
Subsequently, single crystals were grown by a chemical vapor transport process using I${_2}$ (approximately 0.05 g) as the transport agent.
The temperature of the source (sink) was 950\degree C (850\degree C).
Crystals with lateral dimensions up to 2 mm were obtained after being kept at the above temperatures for about one week.
The phase purity of the crystals was checked by powder X-ray diffraction, and the orientation of the single crystalline samples was checked by a Laue diffractometer.
Energy Dispersive X-ray spectroscopy was used to characterize the resulting stoichiometry of the single crystals.
It was measured on the ZEISS 1550VP field emission SEM with Oxford X-Max SDD X-ray Energy Dispersive Spectrometer (EDS) system.
Magnetization was measured by using a Vibrating-sample magnetometer (VSM) option implemented on a commercial superconducting magnet (Quantum Design DynaCool PPMS).

\subsection{Transport Measurement under Strain}
Electrical transport properties were measured using both standard four-(six-) wire methods and modified Montgomery method (See dedicated section below) in commercial cryostats with a superconducting magnet. 
Strain was applied to single crystals of \ce{CoTa3S6} using a commercial piezoelectric uniaxial strain cell (CS-100, Razorbill Instruments). Polished bar-shape samples with typical dimension $1.5\times0.5\times 0.1$ mm$^3$ were mounted to the strain cell with Stycast 2850FT epoxy. For data shown in Fig. 2, the center part of the sample between the mounting plates is approximately 0.7 mm -- we assume this to be the length $L$ of the strained region of the sample and it is used to estimate the applied strain through $\epsilon_{xx}=\Delta L/L$ where $\Delta L$ is obtained from the capacitive displacement sensor in the strain cell. 
A home-made probe was used to load the stain cell in the cryostat. The piezoelectric stacks were controlled by a combination of the DC output of Keithley 6221 current source and TEGAM 2350 high voltage amplifier.  Longitudinal and Hall resistivities were measured using external lock-in amplifiers (Stanford Research Systems, SR860) with an AC current of typical amplitude 2 mA from a voltage controlled current source (Stanford Research Systems, CS580).

\subsection{Modified Montgomery Method}
The modified Montgomery method is an effective technique for measuring the anisotropic resistivity in orthorgonal directions within a single sample simultaneously \cite{montgomery1971method,dos2011procedure,liu2024absence}. 
In this study, \ce{CoTa3S6} is polished into a thin, nearly square shape with its edges aligned along the  $x$- and  $y$-axes, respectively (see Fig. S4 inset for a photo of the Montgomery sample). 
Four electrodes were attached at the corners of the sample with relative small contact area to ensure precise measurements.
The resistance $R_{xx}$  ($R_{yy}$) was determined by applying current $I_x$ ($I_y$) along the $x$-($y$-) axis on one side and measuring the voltage drop $V_x$ ($V_y$) on the opposite side; and $R_{xx}=V_x/I_x$ ($R_{yy}=V_y/I_y$).
The resistivities $\rho_{xx}$  and  $\rho_{yy}$ were calculated based on the ratio of  $R_{xx}$  and  $R_{yy}$  (see the Supplementary Information for details). $\rho_{xx} (\rho_{yy})$ presented in Figs.\ref{fig:figure1},\ref{fig:figure3} and \ref{fig:figure4} are obtained using this method.

\subsection{Polarization rotation measurement}

The schematic of the optical set up can be found in SI Sec. II. A linearly polarized beam is generated by passing light through a polarizer and a polarization-independent beam splitter (half mirror), which transmits half of the beam through a pinhole and discards the other half. The transmitted beam then passes through a half-wave plate (HWP), mechanically rotated such that its fast axis is at an angle $\alpha/2$ to the initial polarization direction. This results in a polarization rotation of angle $\alpha$.

The beam passes through the optical window and reflects from the sample. Upon return, it passes through the HWP again, inducing an additional rotation by $-\alpha$, leaving the polarization direction rotated by $\theta_T$ in total. The beam then encounters the beam splitter again, which directs part of the light to a Wollaston prism. The Wollaston prism, rotated at $\pi/4$, separates the beam into two orthogonal polarization components ($s$ and $p$), which are detected by two balanced detectors ($P1$ and $P2$). For a gold mirror calibration sample, $P1$ and $P2$ are balanced, i.e., $\Delta P = P_1 - P_2 \approx 0$.

The optical field amplitudes at the two detectors are:
\begin{equation}
E_1 = E_0 \cos\left(\frac{\pi}{4} - \theta_T\right), \quad
E_2 = E_0 \cos\left(\frac{\pi}{4} + \theta_T\right)
\end{equation}

Let $E_0$ be the total field amplitude. The total and differential intensities are:
\begin{align}
I_1 + I_2 &= E_1^2 + E_2^2 \nonumber
         = E_0^2\\
I_1 - I_2 &= E_1^2 - E_2^2= E_0^2 \sin(2\theta_T)
\end{align}

Therefore, the normalized signal becomes:
\begin{equation}
\frac{I_1 - I_2}{I_1 + I_2} = \sin(2\theta_T)
\end{equation}

Since optical power $P \propto I$, we can write:
\begin{align}
\theta_T = \frac{1}{2} \arcsin\left(\frac{I_1 - I_2}{I_1 + I_2}\right)
&= \frac{1}{2} \arcsin\left(\frac{P_1 - P_2}{P_1 + P_2}\right)\nonumber\\
&= \frac{1}{2} \arcsin\left(\frac{\Delta P}{P_1 + P_2}\right)
\end{align}
As the $\theta_T$ is a function of $\alpha$, the polar plot can be fit with $\theta_T=\theta_P\sin[2(\alpha-\alpha_0)]$.
$\alpha_0$ defines the orientation of the slow axis, which corresponds to the direction of the nematic director.

\subsection{Comparing in-plane strain and magnetic fields as domain polarization forces}

Due to a linear coupling between in-plane strain and the nematic order, an orthorhombic distortion proportional to the nematic order parameter (e.g. $\epsilon_{xx}-\epsilon_{yy}$ for domain A) is expected for each domain. As a result, in-plane strain-dependence of the nematic order parameter and resistivity anisotropy should be non-hysteretic, and a linear proportionality is expected between the two in a multi-domain state \cite{Bartlett2021-hv}. In contrast, when using the conjugate thermodynamic variable to strain, in-plane uniaxial stress (e.g. $\sigma_{xx}-\sigma_{yy}$) to tune the nematic domains, one may exhibit hysteresis analogous to $M(H)$ curves in ferromagnets. The thermodynamic condition in our experiment is such that strain $\epsilon_{xx}$ is the control parameter, therefore assessing a potentially nonvolatile stress-tuning of the nematic domains is beyond the present experiments. Meanwhile, in-plane magnetic field, itself unconstrained by the nematic order parameters, can exhibit memory effects in the polarization process.

\subsection{Resistivity anisotropy under out-of-plane field}

The magnetic field dependence of the resistivity anisotropy for $H\parallel z$ in Fig. \ref{fig:figure4}(a) was measured in the following sequence.
The sample was mounted on a probe with rotation capability and field-cooled by $H\parallel x$ at 9 T from $T = 50$ K to 2 K, and then the field was set to zero.
After this, the sample platform is slowly rotated to the $H\parallel z$ configuration at fixed temperature.
During this, the effect of magnetic field should be negligible.
The system is then warmed up to the target temperature; 
after waiting for temperature stability, an out-of-plane field is applied from 0 to 9 T where $\rho_{xx}$ and $\rho_{yy}$ are measured simultaneously.

\subsection{Phenomenological modeling of phases in \ce{CoTa3S6}}

The phenomenological free energy of the system is given by Eq. (\ref{FE1}) and Eq. (\ref{FE2}). In Fig. \ref{fig:figure4}c, the collinear AFM order parameter is chosen to be $|\ve{L}_i|$ when the other two  $|\ve{L}_j|,j\neq i$ are zero, and set to zero when at least two of the $|\ve{L}_j|,j=1,2,3$ are non-zero. The non-coplanar AFM order parameter is defined as $|(\ve{L}_1\times\ve{L}_2)\cdot\ve{L}_3|$, which is proportional to the scalar spin chirality of the non-coplanar spin configuration. For the nematic order parameter, we are plotting $n$.
The values of the parameters are as follows: for the nematic order, $b = \tilde{b}(T-T_{\text{n}}), \tilde{b}=0.3,c=-0.05, w=5$, and $T_{\text{n}}=30$; for the magnetic orders, $v_0=5, \tilde{a}=1,\tilde{v}_1=1$, and $T_{N1}=37,T_{N2}=25$. For the coupling between the magnetic orders and nematic order, we take $g=0.05$.

\section{Author Contributions}

Z.F. grew the single crystals and conducted transport and magnetization measurements. Z.F. performed strain experiments with T.L. and J.R.S.. W.L. and J.X. carried out optical birefringence measurements. F.L. and T.L. carried out numerical simulations with M.Y. and L.Y.. The experiments were designed, and the results were analyzed through discussions among Z.F., T.K., and L.Y.. Z.F., T.K., and L.Y. wrote the manuscript with input from all authors.

\section{Data availability}
The datasets generated during and/or analyzed during the current study will be available in the Caltech Research Data Repository when the manuscript is published.
\section{Competing interests}
The authors declare no competing interests.

\bibliography{CoTa3S6}

\begin{thebibliography}{70}%
\makeatletter
\providecommand \@ifxundefined [1]{%
 \@ifx{#1\undefined}
}%
\providecommand \@ifnum [1]{%
 \ifnum #1\expandafter \@firstoftwo
 \else \expandafter \@secondoftwo
 \fi
}%
\providecommand \@ifx [1]{%
 \ifx #1\expandafter \@firstoftwo
 \else \expandafter \@secondoftwo
 \fi
}%
\providecommand \natexlab [1]{#1}%
\providecommand \enquote  [1]{``#1''}%
\providecommand \bibnamefont  [1]{#1}%
\providecommand \bibfnamefont [1]{#1}%
\providecommand \citenamefont [1]{#1}%
\providecommand \href@noop [0]{\@secondoftwo}%
\providecommand \href [0]{\begingroup \@sanitize@url \@href}%
\providecommand \@href[1]{\@@startlink{#1}\@@href}%
\providecommand \@@href[1]{\endgroup#1\@@endlink}%
\providecommand \@sanitize@url [0]{\catcode `\\12\catcode `\$12\catcode `\&12\catcode `\#12\catcode `\^12\catcode `\_12\catcode `\%12\relax}%
\providecommand \@@startlink[1]{}%
\providecommand \@@endlink[0]{}%
\providecommand \url  [0]{\begingroup\@sanitize@url \@url }%
\providecommand \@url [1]{\endgroup\@href {#1}{\urlprefix }}%
\providecommand \urlprefix  [0]{URL }%
\providecommand \Eprint [0]{\href }%
\providecommand \doibase [0]{https://doi.org/}%
\providecommand \selectlanguage [0]{\@gobble}%
\providecommand \bibinfo  [0]{\@secondoftwo}%
\providecommand \bibfield  [0]{\@secondoftwo}%
\providecommand \translation [1]{[#1]}%
\providecommand \BibitemOpen [0]{}%
\providecommand \bibitemStop [0]{}%
\providecommand \bibitemNoStop [0]{.\EOS\space}%
\providecommand \EOS [0]{\spacefactor3000\relax}%
\providecommand \BibitemShut  [1]{\csname bibitem#1\endcsname}%
\let\auto@bib@innerbib\@empty
\bibitem [{\citenamefont {Baltz}\ \emph {et~al.}(2018)\citenamefont {Baltz}, \citenamefont {Manchon}, \citenamefont {Tsoi}, \citenamefont {Moriyama}, \citenamefont {Ono},\ and\ \citenamefont {Tserkovnyak}}]{baltz2018antiferromagnetic}%
  \BibitemOpen
  \bibfield  {author} {\bibinfo {author} {\bibfnamefont {V.}~\bibnamefont {Baltz}}, \bibinfo {author} {\bibfnamefont {A.}~\bibnamefont {Manchon}}, \bibinfo {author} {\bibfnamefont {M.}~\bibnamefont {Tsoi}}, \bibinfo {author} {\bibfnamefont {T.}~\bibnamefont {Moriyama}}, \bibinfo {author} {\bibfnamefont {T.}~\bibnamefont {Ono}},\ and\ \bibinfo {author} {\bibfnamefont {Y.}~\bibnamefont {Tserkovnyak}},\ }\bibfield  {title} {\bibinfo {title} {Antiferromagnetic spintronics},\ }\href@noop {} {\bibfield  {journal} {\bibinfo  {journal} {Rev. Mod. Phys.}\ }\textbf {\bibinfo {volume} {90}},\ \bibinfo {pages} {015005} (\bibinfo {year} {2018})}\BibitemShut {NoStop}%
\bibitem [{\citenamefont {Jungwirth}\ \emph {et~al.}(2016)\citenamefont {Jungwirth}, \citenamefont {Marti}, \citenamefont {Wadley},\ and\ \citenamefont {Wunderlich}}]{jungwirth2016antiferromagnetic}%
  \BibitemOpen
  \bibfield  {author} {\bibinfo {author} {\bibfnamefont {T.}~\bibnamefont {Jungwirth}}, \bibinfo {author} {\bibfnamefont {X.}~\bibnamefont {Marti}}, \bibinfo {author} {\bibfnamefont {P.}~\bibnamefont {Wadley}},\ and\ \bibinfo {author} {\bibfnamefont {J.}~\bibnamefont {Wunderlich}},\ }\bibfield  {title} {\bibinfo {title} {Antiferromagnetic spintronics},\ }\href@noop {} {\bibfield  {journal} {\bibinfo  {journal} {Nature nanotechnology}\ }\textbf {\bibinfo {volume} {11}},\ \bibinfo {pages} {231} (\bibinfo {year} {2016})}\BibitemShut {NoStop}%
\bibitem [{\citenamefont {{\v{S}}mejkal}\ \emph {et~al.}(2018)\citenamefont {{\v{S}}mejkal}, \citenamefont {Mokrousov}, \citenamefont {Yan},\ and\ \citenamefont {MacDonald}}]{vsmejkal2018topological}%
  \BibitemOpen
  \bibfield  {author} {\bibinfo {author} {\bibfnamefont {L.}~\bibnamefont {{\v{S}}mejkal}}, \bibinfo {author} {\bibfnamefont {Y.}~\bibnamefont {Mokrousov}}, \bibinfo {author} {\bibfnamefont {B.}~\bibnamefont {Yan}},\ and\ \bibinfo {author} {\bibfnamefont {A.~H.}\ \bibnamefont {MacDonald}},\ }\bibfield  {title} {\bibinfo {title} {Topological antiferromagnetic spintronics},\ }\href@noop {} {\bibfield  {journal} {\bibinfo  {journal} {Nature physics}\ }\textbf {\bibinfo {volume} {14}},\ \bibinfo {pages} {242} (\bibinfo {year} {2018})}\BibitemShut {NoStop}%
\bibitem [{\citenamefont {Marti}\ \emph {et~al.}(2014)\citenamefont {Marti}, \citenamefont {Fina}, \citenamefont {Frontera}, \citenamefont {Liu}, \citenamefont {Wadley}, \citenamefont {He}, \citenamefont {Paull}, \citenamefont {Clarkson}, \citenamefont {Kudrnovský}, \citenamefont {Turek}, \citenamefont {Kuneš}, \citenamefont {Yi}, \citenamefont {Chu}, \citenamefont {Nelson}, \citenamefont {You}, \citenamefont {Arenholz}, \citenamefont {Salahuddin}, \citenamefont {Fontcuberta}, \citenamefont {Jungwirth},\ and\ \citenamefont {Ramesh}}]{Marti2014-mx}%
  \BibitemOpen
  \bibfield  {author} {\bibinfo {author} {\bibfnamefont {X.}~\bibnamefont {Marti}}, \bibinfo {author} {\bibfnamefont {I.}~\bibnamefont {Fina}}, \bibinfo {author} {\bibfnamefont {C.}~\bibnamefont {Frontera}}, \bibinfo {author} {\bibfnamefont {J.}~\bibnamefont {Liu}}, \bibinfo {author} {\bibfnamefont {P.}~\bibnamefont {Wadley}}, \bibinfo {author} {\bibfnamefont {Q.}~\bibnamefont {He}}, \bibinfo {author} {\bibfnamefont {R.~J.}\ \bibnamefont {Paull}}, \bibinfo {author} {\bibfnamefont {J.~D.}\ \bibnamefont {Clarkson}}, \bibinfo {author} {\bibfnamefont {J.}~\bibnamefont {Kudrnovský}}, \bibinfo {author} {\bibfnamefont {I.}~\bibnamefont {Turek}}, \bibinfo {author} {\bibfnamefont {J.}~\bibnamefont {Kuneš}}, \bibinfo {author} {\bibfnamefont {D.}~\bibnamefont {Yi}}, \bibinfo {author} {\bibfnamefont {J.-H.}\ \bibnamefont {Chu}}, \bibinfo {author} {\bibfnamefont {C.~T.}\ \bibnamefont {Nelson}}, \bibinfo {author} {\bibfnamefont {L.}~\bibnamefont {You}}, \bibinfo {author} {\bibfnamefont {E.}~\bibnamefont {Arenholz}},
  \bibinfo {author} {\bibfnamefont {S.}~\bibnamefont {Salahuddin}}, \bibinfo {author} {\bibfnamefont {J.}~\bibnamefont {Fontcuberta}}, \bibinfo {author} {\bibfnamefont {T.}~\bibnamefont {Jungwirth}},\ and\ \bibinfo {author} {\bibfnamefont {R.}~\bibnamefont {Ramesh}},\ }\bibfield  {title} {\bibinfo {title} {Room-temperature antiferromagnetic memory resistor},\ }\href@noop {} {\bibfield  {journal} {\bibinfo  {journal} {Nat. Mater.}\ }\textbf {\bibinfo {volume} {13}},\ \bibinfo {pages} {367} (\bibinfo {year} {2014})}\BibitemShut {NoStop}%
\bibitem [{\citenamefont {Wadley}\ \emph {et~al.}(2016)\citenamefont {Wadley}, \citenamefont {Howells}, \citenamefont {Železný}, \citenamefont {Andrews}, \citenamefont {Hills}, \citenamefont {Campion}, \citenamefont {Novák}, \citenamefont {Olejník}, \citenamefont {Maccherozzi}, \citenamefont {Dhesi}, \citenamefont {Martin}, \citenamefont {Wagner}, \citenamefont {Wunderlich}, \citenamefont {Freimuth}, \citenamefont {Mokrousov}, \citenamefont {Kuneš}, \citenamefont {Chauhan}, \citenamefont {Grzybowski}, \citenamefont {Rushforth}, \citenamefont {Edmonds}, \citenamefont {Gallagher},\ and\ \citenamefont {Jungwirth}}]{Wadley2016-fb}%
  \BibitemOpen
  \bibfield  {author} {\bibinfo {author} {\bibfnamefont {P.}~\bibnamefont {Wadley}}, \bibinfo {author} {\bibfnamefont {B.}~\bibnamefont {Howells}}, \bibinfo {author} {\bibfnamefont {J.}~\bibnamefont {Železný}}, \bibinfo {author} {\bibfnamefont {C.}~\bibnamefont {Andrews}}, \bibinfo {author} {\bibfnamefont {V.}~\bibnamefont {Hills}}, \bibinfo {author} {\bibfnamefont {R.~P.}\ \bibnamefont {Campion}}, \bibinfo {author} {\bibfnamefont {V.}~\bibnamefont {Novák}}, \bibinfo {author} {\bibfnamefont {K.}~\bibnamefont {Olejník}}, \bibinfo {author} {\bibfnamefont {F.}~\bibnamefont {Maccherozzi}}, \bibinfo {author} {\bibfnamefont {S.~S.}\ \bibnamefont {Dhesi}}, \bibinfo {author} {\bibfnamefont {S.~Y.}\ \bibnamefont {Martin}}, \bibinfo {author} {\bibfnamefont {T.}~\bibnamefont {Wagner}}, \bibinfo {author} {\bibfnamefont {J.}~\bibnamefont {Wunderlich}}, \bibinfo {author} {\bibfnamefont {F.}~\bibnamefont {Freimuth}}, \bibinfo {author} {\bibfnamefont {Y.}~\bibnamefont {Mokrousov}}, \bibinfo {author} {\bibfnamefont
  {J.}~\bibnamefont {Kuneš}}, \bibinfo {author} {\bibfnamefont {J.~S.}\ \bibnamefont {Chauhan}}, \bibinfo {author} {\bibfnamefont {M.~J.}\ \bibnamefont {Grzybowski}}, \bibinfo {author} {\bibfnamefont {A.~W.}\ \bibnamefont {Rushforth}}, \bibinfo {author} {\bibfnamefont {K.~W.}\ \bibnamefont {Edmonds}}, \bibinfo {author} {\bibfnamefont {B.~L.}\ \bibnamefont {Gallagher}},\ and\ \bibinfo {author} {\bibfnamefont {T.}~\bibnamefont {Jungwirth}},\ }\bibfield  {title} {\bibinfo {title} {Electrical switching of an antiferromagnet},\ }\href@noop {} {\bibfield  {journal} {\bibinfo  {journal} {Science}\ }\textbf {\bibinfo {volume} {351}},\ \bibinfo {pages} {587} (\bibinfo {year} {2016})}\BibitemShut {NoStop}%
\bibitem [{\citenamefont {Železný}\ \emph {et~al.}(2014)\citenamefont {Železný}, \citenamefont {Gao}, \citenamefont {Výborný}, \citenamefont {Zemen}, \citenamefont {Mašek}, \citenamefont {Manchon}, \citenamefont {Wunderlich}, \citenamefont {Sinova},\ and\ \citenamefont {Jungwirth}}]{Zelezny2014-zx}%
  \BibitemOpen
  \bibfield  {author} {\bibinfo {author} {\bibfnamefont {J.}~\bibnamefont {Železný}}, \bibinfo {author} {\bibfnamefont {H.}~\bibnamefont {Gao}}, \bibinfo {author} {\bibfnamefont {K.}~\bibnamefont {Výborný}}, \bibinfo {author} {\bibfnamefont {J.}~\bibnamefont {Zemen}}, \bibinfo {author} {\bibfnamefont {J.}~\bibnamefont {Mašek}}, \bibinfo {author} {\bibfnamefont {A.}~\bibnamefont {Manchon}}, \bibinfo {author} {\bibfnamefont {J.}~\bibnamefont {Wunderlich}}, \bibinfo {author} {\bibfnamefont {J.}~\bibnamefont {Sinova}},\ and\ \bibinfo {author} {\bibfnamefont {T.}~\bibnamefont {Jungwirth}},\ }\bibfield  {title} {\bibinfo {title} {Relativistic néel-order fields induced by electrical current in antiferromagnets},\ }\href@noop {} {\bibfield  {journal} {\bibinfo  {journal} {Phys. Rev. Lett.}\ }\textbf {\bibinfo {volume} {113}},\ \bibinfo {pages} {157201} (\bibinfo {year} {2014})}\BibitemShut {NoStop}%
\bibitem [{\citenamefont {Grzybowski}\ \emph {et~al.}(2017)\citenamefont {Grzybowski}, \citenamefont {Wadley}, \citenamefont {Edmonds}, \citenamefont {Beardsley}, \citenamefont {Hills}, \citenamefont {Campion}, \citenamefont {Gallagher}, \citenamefont {Chauhan}, \citenamefont {Novak}, \citenamefont {Jungwirth}, \citenamefont {Maccherozzi},\ and\ \citenamefont {Dhesi}}]{Grzybowski2017-ou}%
  \BibitemOpen
  \bibfield  {author} {\bibinfo {author} {\bibfnamefont {M.~J.}\ \bibnamefont {Grzybowski}}, \bibinfo {author} {\bibfnamefont {P.}~\bibnamefont {Wadley}}, \bibinfo {author} {\bibfnamefont {K.~W.}\ \bibnamefont {Edmonds}}, \bibinfo {author} {\bibfnamefont {R.}~\bibnamefont {Beardsley}}, \bibinfo {author} {\bibfnamefont {V.}~\bibnamefont {Hills}}, \bibinfo {author} {\bibfnamefont {R.~P.}\ \bibnamefont {Campion}}, \bibinfo {author} {\bibfnamefont {B.~L.}\ \bibnamefont {Gallagher}}, \bibinfo {author} {\bibfnamefont {J.~S.}\ \bibnamefont {Chauhan}}, \bibinfo {author} {\bibfnamefont {V.}~\bibnamefont {Novak}}, \bibinfo {author} {\bibfnamefont {T.}~\bibnamefont {Jungwirth}}, \bibinfo {author} {\bibfnamefont {F.}~\bibnamefont {Maccherozzi}},\ and\ \bibinfo {author} {\bibfnamefont {S.~S.}\ \bibnamefont {Dhesi}},\ }\bibfield  {title} {\bibinfo {title} {Imaging current-induced switching of antiferromagnetic domains in \ce{CuMnAs}},\ }\href@noop {} {\bibfield  {journal} {\bibinfo  {journal} {Phys. Rev. Lett.}\ }\textbf
  {\bibinfo {volume} {118}},\ \bibinfo {pages} {057701} (\bibinfo {year} {2017})}\BibitemShut {NoStop}%
\bibitem [{\citenamefont {Nair}\ \emph {et~al.}(2020)\citenamefont {Nair}, \citenamefont {Maniv}, \citenamefont {John}, \citenamefont {Doyle}, \citenamefont {Orenstein},\ and\ \citenamefont {Analytis}}]{Nair2020-ts}%
  \BibitemOpen
  \bibfield  {author} {\bibinfo {author} {\bibfnamefont {N.~L.}\ \bibnamefont {Nair}}, \bibinfo {author} {\bibfnamefont {E.}~\bibnamefont {Maniv}}, \bibinfo {author} {\bibfnamefont {C.}~\bibnamefont {John}}, \bibinfo {author} {\bibfnamefont {S.}~\bibnamefont {Doyle}}, \bibinfo {author} {\bibfnamefont {J.}~\bibnamefont {Orenstein}},\ and\ \bibinfo {author} {\bibfnamefont {J.~G.}\ \bibnamefont {Analytis}},\ }\bibfield  {title} {\bibinfo {title} {Electrical switching in a magnetically intercalated transition metal dichalcogenide},\ }\href@noop {} {\bibfield  {journal} {\bibinfo  {journal} {Nat. Mater.}\ }\textbf {\bibinfo {volume} {19}},\ \bibinfo {pages} {153} (\bibinfo {year} {2020})}\BibitemShut {NoStop}%
\bibitem [{\citenamefont {Gao}\ \emph {et~al.}(2023)\citenamefont {Gao}, \citenamefont {Liu}, \citenamefont {Qiu}, \citenamefont {Ghosh}, \citenamefont {Trevisan}, \citenamefont {Onishi}, \citenamefont {Hu}, \citenamefont {Qian}, \citenamefont {Tien}, \citenamefont {Chen}, \citenamefont {Huang}, \citenamefont {Bérubé}, \citenamefont {Li}, \citenamefont {Tzschaschel}, \citenamefont {Dinh}, \citenamefont {Sun}, \citenamefont {Ho}, \citenamefont {Lien}, \citenamefont {Singh}, \citenamefont {Watanabe}, \citenamefont {Taniguchi}, \citenamefont {Bell}, \citenamefont {Lin}, \citenamefont {Chang}, \citenamefont {Du}, \citenamefont {Bansil}, \citenamefont {Fu}, \citenamefont {Ni}, \citenamefont {Orth}, \citenamefont {Ma},\ and\ \citenamefont {Xu}}]{Gao2023-au}%
  \BibitemOpen
  \bibfield  {author} {\bibinfo {author} {\bibfnamefont {A.}~\bibnamefont {Gao}}, \bibinfo {author} {\bibfnamefont {Y.-F.}\ \bibnamefont {Liu}}, \bibinfo {author} {\bibfnamefont {J.-X.}\ \bibnamefont {Qiu}}, \bibinfo {author} {\bibfnamefont {B.}~\bibnamefont {Ghosh}}, \bibinfo {author} {\bibfnamefont {T.~V.}\ \bibnamefont {Trevisan}}, \bibinfo {author} {\bibfnamefont {Y.}~\bibnamefont {Onishi}}, \bibinfo {author} {\bibfnamefont {C.}~\bibnamefont {Hu}}, \bibinfo {author} {\bibfnamefont {T.}~\bibnamefont {Qian}}, \bibinfo {author} {\bibfnamefont {H.-J.}\ \bibnamefont {Tien}}, \bibinfo {author} {\bibfnamefont {S.-W.}\ \bibnamefont {Chen}}, \bibinfo {author} {\bibfnamefont {M.}~\bibnamefont {Huang}}, \bibinfo {author} {\bibfnamefont {D.}~\bibnamefont {Bérubé}}, \bibinfo {author} {\bibfnamefont {H.}~\bibnamefont {Li}}, \bibinfo {author} {\bibfnamefont {C.}~\bibnamefont {Tzschaschel}}, \bibinfo {author} {\bibfnamefont {T.}~\bibnamefont {Dinh}}, \bibinfo {author} {\bibfnamefont {Z.}~\bibnamefont {Sun}}, \bibinfo
  {author} {\bibfnamefont {S.-C.}\ \bibnamefont {Ho}}, \bibinfo {author} {\bibfnamefont {S.-W.}\ \bibnamefont {Lien}}, \bibinfo {author} {\bibfnamefont {B.}~\bibnamefont {Singh}}, \bibinfo {author} {\bibfnamefont {K.}~\bibnamefont {Watanabe}}, \bibinfo {author} {\bibfnamefont {T.}~\bibnamefont {Taniguchi}}, \bibinfo {author} {\bibfnamefont {D.~C.}\ \bibnamefont {Bell}}, \bibinfo {author} {\bibfnamefont {H.}~\bibnamefont {Lin}}, \bibinfo {author} {\bibfnamefont {T.-R.}\ \bibnamefont {Chang}}, \bibinfo {author} {\bibfnamefont {C.~R.}\ \bibnamefont {Du}}, \bibinfo {author} {\bibfnamefont {A.}~\bibnamefont {Bansil}}, \bibinfo {author} {\bibfnamefont {L.}~\bibnamefont {Fu}}, \bibinfo {author} {\bibfnamefont {N.}~\bibnamefont {Ni}}, \bibinfo {author} {\bibfnamefont {P.~P.}\ \bibnamefont {Orth}}, \bibinfo {author} {\bibfnamefont {Q.}~\bibnamefont {Ma}},\ and\ \bibinfo {author} {\bibfnamefont {S.-Y.}\ \bibnamefont {Xu}},\ }\bibfield  {title} {\bibinfo {title} {Quantum metric nonlinear hall effect in a topological
  antiferromagnetic heterostructure},\ }\href@noop {} {\bibfield  {journal} {\bibinfo  {journal} {Science}\ }\textbf {\bibinfo {volume} {381}},\ \bibinfo {pages} {181} (\bibinfo {year} {2023})}\BibitemShut {NoStop}%
\bibitem [{\citenamefont {Fang}\ \emph {et~al.}(2024)\citenamefont {Fang}, \citenamefont {Cano},\ and\ \citenamefont {Ghorashi}}]{PhysRevLett.133.106701}%
  \BibitemOpen
  \bibfield  {author} {\bibinfo {author} {\bibfnamefont {Y.}~\bibnamefont {Fang}}, \bibinfo {author} {\bibfnamefont {J.}~\bibnamefont {Cano}},\ and\ \bibinfo {author} {\bibfnamefont {S.~A.~A.}\ \bibnamefont {Ghorashi}},\ }\bibfield  {title} {\bibinfo {title} {Quantum geometry induced nonlinear transport in altermagnets},\ }\href@noop {} {\bibfield  {journal} {\bibinfo  {journal} {Phys. Rev. Lett.}\ }\textbf {\bibinfo {volume} {133}},\ \bibinfo {pages} {106701} (\bibinfo {year} {2024})}\BibitemShut {NoStop}%
\bibitem [{\citenamefont {Kivelson}\ \emph {et~al.}(1998)\citenamefont {Kivelson}, \citenamefont {Fradkin},\ and\ \citenamefont {Emery}}]{Kivelson1998-ki}%
  \BibitemOpen
  \bibfield  {author} {\bibinfo {author} {\bibfnamefont {S.~A.}\ \bibnamefont {Kivelson}}, \bibinfo {author} {\bibfnamefont {E.}~\bibnamefont {Fradkin}},\ and\ \bibinfo {author} {\bibfnamefont {V.~J.}\ \bibnamefont {Emery}},\ }\bibfield  {title} {\bibinfo {title} {Electronic liquid-crystal phases of a doped mott insulator},\ }\href@noop {} {\bibfield  {journal} {\bibinfo  {journal} {Nature}\ }\textbf {\bibinfo {volume} {393}},\ \bibinfo {pages} {550} (\bibinfo {year} {1998})}\BibitemShut {NoStop}%
\bibitem [{\citenamefont {Fradkin}\ \emph {et~al.}(2010)\citenamefont {Fradkin}, \citenamefont {Kivelson}, \citenamefont {Lawler}, \citenamefont {Eisenstein},\ and\ \citenamefont {Mackenzie}}]{fradkin2010nematic}%
  \BibitemOpen
  \bibfield  {author} {\bibinfo {author} {\bibfnamefont {E.}~\bibnamefont {Fradkin}}, \bibinfo {author} {\bibfnamefont {S.~A.}\ \bibnamefont {Kivelson}}, \bibinfo {author} {\bibfnamefont {M.~J.}\ \bibnamefont {Lawler}}, \bibinfo {author} {\bibfnamefont {J.~P.}\ \bibnamefont {Eisenstein}},\ and\ \bibinfo {author} {\bibfnamefont {A.~P.}\ \bibnamefont {Mackenzie}},\ }\bibfield  {title} {\bibinfo {title} {Nematic fermi fluids in condensed matter physics},\ }\href@noop {} {\bibfield  {journal} {\bibinfo  {journal} {Annu. Rev. Condens. Matter Phys.}\ }\textbf {\bibinfo {volume} {1}},\ \bibinfo {pages} {153} (\bibinfo {year} {2010})}\BibitemShut {NoStop}%
\bibitem [{\citenamefont {Lilly}\ \emph {et~al.}(1999{\natexlab{a}})\citenamefont {Lilly}, \citenamefont {Cooper}, \citenamefont {Eisenstein}, \citenamefont {Pfeiffer},\ and\ \citenamefont {West}}]{lilly1999evidence}%
  \BibitemOpen
  \bibfield  {author} {\bibinfo {author} {\bibfnamefont {M.}~\bibnamefont {Lilly}}, \bibinfo {author} {\bibfnamefont {K.}~\bibnamefont {Cooper}}, \bibinfo {author} {\bibfnamefont {J.}~\bibnamefont {Eisenstein}}, \bibinfo {author} {\bibfnamefont {L.}~\bibnamefont {Pfeiffer}},\ and\ \bibinfo {author} {\bibfnamefont {K.}~\bibnamefont {West}},\ }\bibfield  {title} {\bibinfo {title} {Evidence for an anisotropic state of two-dimensional electrons in high landau levels},\ }\href@noop {} {\bibfield  {journal} {\bibinfo  {journal} {Phys. Rev. Lett.}\ }\textbf {\bibinfo {volume} {82}},\ \bibinfo {pages} {394} (\bibinfo {year} {1999}{\natexlab{a}})}\BibitemShut {NoStop}%
\bibitem [{\citenamefont {Borzi}\ \emph {et~al.}(2007)\citenamefont {Borzi}, \citenamefont {Grigera}, \citenamefont {Farrell}, \citenamefont {Perry}, \citenamefont {Lister}, \citenamefont {Lee}, \citenamefont {Tennant}, \citenamefont {Maeno},\ and\ \citenamefont {Mackenzie}}]{Borzi2007-li}%
  \BibitemOpen
  \bibfield  {author} {\bibinfo {author} {\bibfnamefont {R.~A.}\ \bibnamefont {Borzi}}, \bibinfo {author} {\bibfnamefont {S.~A.}\ \bibnamefont {Grigera}}, \bibinfo {author} {\bibfnamefont {J.}~\bibnamefont {Farrell}}, \bibinfo {author} {\bibfnamefont {R.~S.}\ \bibnamefont {Perry}}, \bibinfo {author} {\bibfnamefont {S.~J.~S.}\ \bibnamefont {Lister}}, \bibinfo {author} {\bibfnamefont {S.~L.}\ \bibnamefont {Lee}}, \bibinfo {author} {\bibfnamefont {D.~A.}\ \bibnamefont {Tennant}}, \bibinfo {author} {\bibfnamefont {Y.}~\bibnamefont {Maeno}},\ and\ \bibinfo {author} {\bibfnamefont {A.~P.}\ \bibnamefont {Mackenzie}},\ }\bibfield  {title} {\bibinfo {title} {Formation of a nematic fluid at high fields in \ce{Sr3Ru2O7}},\ }\href@noop {} {\bibfield  {journal} {\bibinfo  {journal} {Science}\ }\textbf {\bibinfo {volume} {315}},\ \bibinfo {pages} {214} (\bibinfo {year} {2007})}\BibitemShut {NoStop}%
\bibitem [{\citenamefont {Hinkov}\ \emph {et~al.}(2008)\citenamefont {Hinkov}, \citenamefont {Haug}, \citenamefont {Fauqué}, \citenamefont {Bourges}, \citenamefont {Sidis}, \citenamefont {Ivanov}, \citenamefont {Bernhard}, \citenamefont {Lin},\ and\ \citenamefont {Keimer}}]{Hinkov2008-dh}%
  \BibitemOpen
  \bibfield  {author} {\bibinfo {author} {\bibfnamefont {V.}~\bibnamefont {Hinkov}}, \bibinfo {author} {\bibfnamefont {D.}~\bibnamefont {Haug}}, \bibinfo {author} {\bibfnamefont {B.}~\bibnamefont {Fauqué}}, \bibinfo {author} {\bibfnamefont {P.}~\bibnamefont {Bourges}}, \bibinfo {author} {\bibfnamefont {Y.}~\bibnamefont {Sidis}}, \bibinfo {author} {\bibfnamefont {A.}~\bibnamefont {Ivanov}}, \bibinfo {author} {\bibfnamefont {C.}~\bibnamefont {Bernhard}}, \bibinfo {author} {\bibfnamefont {C.~T.}\ \bibnamefont {Lin}},\ and\ \bibinfo {author} {\bibfnamefont {B.}~\bibnamefont {Keimer}},\ }\bibfield  {title} {\bibinfo {title} {Electronic liquid crystal state in the high-temperature superconductor \ce{YBa2Cu3O6.45}},\ }\href@noop {} {\bibfield  {journal} {\bibinfo  {journal} {Science}\ }\textbf {\bibinfo {volume} {319}},\ \bibinfo {pages} {597} (\bibinfo {year} {2008})}\BibitemShut {NoStop}%
\bibitem [{\citenamefont {Chu}\ \emph {et~al.}(2010)\citenamefont {Chu}, \citenamefont {Analytis}, \citenamefont {De~Greve}, \citenamefont {McMahon}, \citenamefont {Islam}, \citenamefont {Yamamoto},\ and\ \citenamefont {Fisher}}]{Chu2010-cl}%
  \BibitemOpen
  \bibfield  {author} {\bibinfo {author} {\bibfnamefont {J.-H.}\ \bibnamefont {Chu}}, \bibinfo {author} {\bibfnamefont {J.~G.}\ \bibnamefont {Analytis}}, \bibinfo {author} {\bibfnamefont {K.}~\bibnamefont {De~Greve}}, \bibinfo {author} {\bibfnamefont {P.~L.}\ \bibnamefont {McMahon}}, \bibinfo {author} {\bibfnamefont {Z.}~\bibnamefont {Islam}}, \bibinfo {author} {\bibfnamefont {Y.}~\bibnamefont {Yamamoto}},\ and\ \bibinfo {author} {\bibfnamefont {I.~R.}\ \bibnamefont {Fisher}},\ }\bibfield  {title} {\bibinfo {title} {In-plane resistivity anisotropy in an underdoped iron arsenide superconductor},\ }\href@noop {} {\bibfield  {journal} {\bibinfo  {journal} {Science}\ }\textbf {\bibinfo {volume} {329}},\ \bibinfo {pages} {824} (\bibinfo {year} {2010})}\BibitemShut {NoStop}%
\bibitem [{\citenamefont {B{\"o}hmer}\ \emph {et~al.}(2022)\citenamefont {B{\"o}hmer}, \citenamefont {Chu}, \citenamefont {Lederer},\ and\ \citenamefont {Yi}}]{bohmer2022nematicity}%
  \BibitemOpen
  \bibfield  {author} {\bibinfo {author} {\bibfnamefont {A.~E.}\ \bibnamefont {B{\"o}hmer}}, \bibinfo {author} {\bibfnamefont {J.-H.}\ \bibnamefont {Chu}}, \bibinfo {author} {\bibfnamefont {S.}~\bibnamefont {Lederer}},\ and\ \bibinfo {author} {\bibfnamefont {M.}~\bibnamefont {Yi}},\ }\bibfield  {title} {\bibinfo {title} {Nematicity and nematic fluctuations in iron-based superconductors},\ }\href@noop {} {\bibfield  {journal} {\bibinfo  {journal} {Nature Physics}\ }\textbf {\bibinfo {volume} {18}},\ \bibinfo {pages} {1412} (\bibinfo {year} {2022})}\BibitemShut {NoStop}%
\bibitem [{\citenamefont {Fernandes}\ \emph {et~al.}(2022)\citenamefont {Fernandes}, \citenamefont {Coldea}, \citenamefont {Ding}, \citenamefont {Fisher}, \citenamefont {Hirschfeld},\ and\ \citenamefont {Kotliar}}]{Fernandes2022-vl}%
  \BibitemOpen
  \bibfield  {author} {\bibinfo {author} {\bibfnamefont {R.~M.}\ \bibnamefont {Fernandes}}, \bibinfo {author} {\bibfnamefont {A.~I.}\ \bibnamefont {Coldea}}, \bibinfo {author} {\bibfnamefont {H.}~\bibnamefont {Ding}}, \bibinfo {author} {\bibfnamefont {I.~R.}\ \bibnamefont {Fisher}}, \bibinfo {author} {\bibfnamefont {P.~J.}\ \bibnamefont {Hirschfeld}},\ and\ \bibinfo {author} {\bibfnamefont {G.}~\bibnamefont {Kotliar}},\ }\bibfield  {title} {\bibinfo {title} {Iron pnictides and chalcogenides: a new paradigm for superconductivity},\ }\href@noop {} {\bibfield  {journal} {\bibinfo  {journal} {Nature}\ }\textbf {\bibinfo {volume} {601}},\ \bibinfo {pages} {35} (\bibinfo {year} {2022})}\BibitemShut {NoStop}%
\bibitem [{\citenamefont {Cao}\ \emph {et~al.}(2021)\citenamefont {Cao}, \citenamefont {Rodan-Legrain}, \citenamefont {Park}, \citenamefont {Yuan}, \citenamefont {Watanabe}, \citenamefont {Taniguchi}, \citenamefont {Fernandes}, \citenamefont {Fu},\ and\ \citenamefont {Jarillo-Herrero}}]{Cao2021-sv}%
  \BibitemOpen
  \bibfield  {author} {\bibinfo {author} {\bibfnamefont {Y.}~\bibnamefont {Cao}}, \bibinfo {author} {\bibfnamefont {D.}~\bibnamefont {Rodan-Legrain}}, \bibinfo {author} {\bibfnamefont {J.~M.}\ \bibnamefont {Park}}, \bibinfo {author} {\bibfnamefont {N.~F.~Q.}\ \bibnamefont {Yuan}}, \bibinfo {author} {\bibfnamefont {K.}~\bibnamefont {Watanabe}}, \bibinfo {author} {\bibfnamefont {T.}~\bibnamefont {Taniguchi}}, \bibinfo {author} {\bibfnamefont {R.~M.}\ \bibnamefont {Fernandes}}, \bibinfo {author} {\bibfnamefont {L.}~\bibnamefont {Fu}},\ and\ \bibinfo {author} {\bibfnamefont {P.}~\bibnamefont {Jarillo-Herrero}},\ }\bibfield  {title} {\bibinfo {title} {Nematicity and competing orders in superconducting magic-angle graphene},\ }\href@noop {} {\bibfield  {journal} {\bibinfo  {journal} {Science}\ }\textbf {\bibinfo {volume} {372}},\ \bibinfo {pages} {264} (\bibinfo {year} {2021})}\BibitemShut {NoStop}%
\bibitem [{\citenamefont {Rubio-Verdú}\ \emph {et~al.}(2022)\citenamefont {Rubio-Verdú}, \citenamefont {Turkel}, \citenamefont {Song}, \citenamefont {Klebl}, \citenamefont {Samajdar}, \citenamefont {Scheurer}, \citenamefont {Venderbos}, \citenamefont {Watanabe}, \citenamefont {Taniguchi}, \citenamefont {Ochoa}, \citenamefont {Xian}, \citenamefont {Kennes}, \citenamefont {Fernandes}, \citenamefont {Rubio},\ and\ \citenamefont {Pasupathy}}]{Rubio-Verdu2022-tm}%
  \BibitemOpen
  \bibfield  {author} {\bibinfo {author} {\bibfnamefont {C.}~\bibnamefont {Rubio-Verdú}}, \bibinfo {author} {\bibfnamefont {S.}~\bibnamefont {Turkel}}, \bibinfo {author} {\bibfnamefont {Y.}~\bibnamefont {Song}}, \bibinfo {author} {\bibfnamefont {L.}~\bibnamefont {Klebl}}, \bibinfo {author} {\bibfnamefont {R.}~\bibnamefont {Samajdar}}, \bibinfo {author} {\bibfnamefont {M.~S.}\ \bibnamefont {Scheurer}}, \bibinfo {author} {\bibfnamefont {J.~W.~F.}\ \bibnamefont {Venderbos}}, \bibinfo {author} {\bibfnamefont {K.}~\bibnamefont {Watanabe}}, \bibinfo {author} {\bibfnamefont {T.}~\bibnamefont {Taniguchi}}, \bibinfo {author} {\bibfnamefont {H.}~\bibnamefont {Ochoa}}, \bibinfo {author} {\bibfnamefont {L.}~\bibnamefont {Xian}}, \bibinfo {author} {\bibfnamefont {D.~M.}\ \bibnamefont {Kennes}}, \bibinfo {author} {\bibfnamefont {R.~M.}\ \bibnamefont {Fernandes}}, \bibinfo {author} {\bibfnamefont {A.}~\bibnamefont {Rubio}},\ and\ \bibinfo {author} {\bibfnamefont {A.~N.}\ \bibnamefont {Pasupathy}},\ }\bibfield  {title}
  {\bibinfo {title} {Moiré nematic phase in twisted double bilayer graphene},\ }\href@noop {} {\bibfield  {journal} {\bibinfo  {journal} {Nat. Phys.}\ }\textbf {\bibinfo {volume} {18}},\ \bibinfo {pages} {196} (\bibinfo {year} {2022})}\BibitemShut {NoStop}%
\bibitem [{\citenamefont {Xu}\ \emph {et~al.}(2022)\citenamefont {Xu}, \citenamefont {Ni}, \citenamefont {Liu}, \citenamefont {Ortiz}, \citenamefont {Deng}, \citenamefont {Wilson}, \citenamefont {Yan}, \citenamefont {Balents},\ and\ \citenamefont {Wu}}]{Xu2022-hu}%
  \BibitemOpen
  \bibfield  {author} {\bibinfo {author} {\bibfnamefont {Y.}~\bibnamefont {Xu}}, \bibinfo {author} {\bibfnamefont {Z.}~\bibnamefont {Ni}}, \bibinfo {author} {\bibfnamefont {Y.}~\bibnamefont {Liu}}, \bibinfo {author} {\bibfnamefont {B.~R.}\ \bibnamefont {Ortiz}}, \bibinfo {author} {\bibfnamefont {Q.}~\bibnamefont {Deng}}, \bibinfo {author} {\bibfnamefont {S.~D.}\ \bibnamefont {Wilson}}, \bibinfo {author} {\bibfnamefont {B.}~\bibnamefont {Yan}}, \bibinfo {author} {\bibfnamefont {L.}~\bibnamefont {Balents}},\ and\ \bibinfo {author} {\bibfnamefont {L.}~\bibnamefont {Wu}},\ }\bibfield  {title} {\bibinfo {title} {Three-state nematicity and magneto-optical kerr effect in the charge density waves in kagome superconductors},\ }\href@noop {} {\bibfield  {journal} {\bibinfo  {journal} {Nat. Phys.}\ }\textbf {\bibinfo {volume} {18}},\ \bibinfo {pages} {1470} (\bibinfo {year} {2022})}\BibitemShut {NoStop}%
\bibitem [{\citenamefont {Li}\ \emph {et~al.}(2023)\citenamefont {Li}, \citenamefont {Cheng}, \citenamefont {Ortiz}, \citenamefont {Tan}, \citenamefont {Werhahn}, \citenamefont {Zeng}, \citenamefont {Johrendt}, \citenamefont {Yan}, \citenamefont {Wang}, \citenamefont {Wilson},\ and\ \citenamefont {Zeljkovic}}]{Li2023-ag}%
  \BibitemOpen
  \bibfield  {author} {\bibinfo {author} {\bibfnamefont {H.}~\bibnamefont {Li}}, \bibinfo {author} {\bibfnamefont {S.}~\bibnamefont {Cheng}}, \bibinfo {author} {\bibfnamefont {B.~R.}\ \bibnamefont {Ortiz}}, \bibinfo {author} {\bibfnamefont {H.}~\bibnamefont {Tan}}, \bibinfo {author} {\bibfnamefont {D.}~\bibnamefont {Werhahn}}, \bibinfo {author} {\bibfnamefont {K.}~\bibnamefont {Zeng}}, \bibinfo {author} {\bibfnamefont {D.}~\bibnamefont {Johrendt}}, \bibinfo {author} {\bibfnamefont {B.}~\bibnamefont {Yan}}, \bibinfo {author} {\bibfnamefont {Z.}~\bibnamefont {Wang}}, \bibinfo {author} {\bibfnamefont {S.~D.}\ \bibnamefont {Wilson}},\ and\ \bibinfo {author} {\bibfnamefont {I.}~\bibnamefont {Zeljkovic}},\ }\bibfield  {title} {\bibinfo {title} {Electronic nematicity without charge density waves in titanium-based kagome metal},\ }\href@noop {} {\bibfield  {journal} {\bibinfo  {journal} {Nat. Phys.}\ }\textbf {\bibinfo {volume} {19}},\ \bibinfo {pages} {1591} (\bibinfo {year} {2023})}\BibitemShut {NoStop}%
\bibitem [{\citenamefont {Yang}\ \emph {et~al.}(2024)\citenamefont {Yang}, \citenamefont {Ye}, \citenamefont {Zhao}, \citenamefont {Liu}, \citenamefont {Yi}, \citenamefont {Zhang}, \citenamefont {Xiao}, \citenamefont {Shi}, \citenamefont {You}, \citenamefont {Huang}, \citenamefont {Wang}, \citenamefont {Wang}, \citenamefont {Guo}, \citenamefont {Lin}, \citenamefont {Shen}, \citenamefont {Zhou}, \citenamefont {Chen}, \citenamefont {Dong}, \citenamefont {Su}, \citenamefont {Wang},\ and\ \citenamefont {Gao}}]{Yang2024-kf}%
  \BibitemOpen
  \bibfield  {author} {\bibinfo {author} {\bibfnamefont {H.}~\bibnamefont {Yang}}, \bibinfo {author} {\bibfnamefont {Y.}~\bibnamefont {Ye}}, \bibinfo {author} {\bibfnamefont {Z.}~\bibnamefont {Zhao}}, \bibinfo {author} {\bibfnamefont {J.}~\bibnamefont {Liu}}, \bibinfo {author} {\bibfnamefont {X.-W.}\ \bibnamefont {Yi}}, \bibinfo {author} {\bibfnamefont {Y.}~\bibnamefont {Zhang}}, \bibinfo {author} {\bibfnamefont {H.}~\bibnamefont {Xiao}}, \bibinfo {author} {\bibfnamefont {J.}~\bibnamefont {Shi}}, \bibinfo {author} {\bibfnamefont {J.-Y.}\ \bibnamefont {You}}, \bibinfo {author} {\bibfnamefont {Z.}~\bibnamefont {Huang}}, \bibinfo {author} {\bibfnamefont {B.}~\bibnamefont {Wang}}, \bibinfo {author} {\bibfnamefont {J.}~\bibnamefont {Wang}}, \bibinfo {author} {\bibfnamefont {H.}~\bibnamefont {Guo}}, \bibinfo {author} {\bibfnamefont {X.}~\bibnamefont {Lin}}, \bibinfo {author} {\bibfnamefont {C.}~\bibnamefont {Shen}}, \bibinfo {author} {\bibfnamefont {W.}~\bibnamefont {Zhou}}, \bibinfo {author} {\bibfnamefont
  {H.}~\bibnamefont {Chen}}, \bibinfo {author} {\bibfnamefont {X.}~\bibnamefont {Dong}}, \bibinfo {author} {\bibfnamefont {G.}~\bibnamefont {Su}}, \bibinfo {author} {\bibfnamefont {Z.}~\bibnamefont {Wang}},\ and\ \bibinfo {author} {\bibfnamefont {H.-J.}\ \bibnamefont {Gao}},\ }\bibfield  {title} {\bibinfo {title} {Superconductivity and nematic order in a new titanium-based kagome metal {CsTi3Bi5} without charge density wave order},\ }\href@noop {} {\bibfield  {journal} {\bibinfo  {journal} {Nat. Commun.}\ }\textbf {\bibinfo {volume} {15}},\ \bibinfo {pages} {9626} (\bibinfo {year} {2024})}\BibitemShut {NoStop}%
\bibitem [{\citenamefont {Fernandes}\ and\ \citenamefont {Venderbos}(2020)}]{Fernandes2020-us}%
  \BibitemOpen
  \bibfield  {author} {\bibinfo {author} {\bibfnamefont {R.~M.}\ \bibnamefont {Fernandes}}\ and\ \bibinfo {author} {\bibfnamefont {J.~W.~F.}\ \bibnamefont {Venderbos}},\ }\bibfield  {title} {\bibinfo {title} {Nematicity with a twist: Rotational symmetry breaking in a moiré superlattice},\ }\href@noop {} {\bibfield  {journal} {\bibinfo  {journal} {Sci. Adv.}\ }\textbf {\bibinfo {volume} {6}},\ \bibinfo {pages} {eaba8834} (\bibinfo {year} {2020})}\BibitemShut {NoStop}%
\bibitem [{\citenamefont {Little}\ \emph {et~al.}(2020)\citenamefont {Little}, \citenamefont {Lee}, \citenamefont {John}, \citenamefont {Doyle}, \citenamefont {Maniv}, \citenamefont {Nair}, \citenamefont {Chen}, \citenamefont {Rees}, \citenamefont {Venderbos}, \citenamefont {Fernandes} \emph {et~al.}}]{little2020three}%
  \BibitemOpen
  \bibfield  {author} {\bibinfo {author} {\bibfnamefont {A.}~\bibnamefont {Little}}, \bibinfo {author} {\bibfnamefont {C.}~\bibnamefont {Lee}}, \bibinfo {author} {\bibfnamefont {C.}~\bibnamefont {John}}, \bibinfo {author} {\bibfnamefont {S.}~\bibnamefont {Doyle}}, \bibinfo {author} {\bibfnamefont {E.}~\bibnamefont {Maniv}}, \bibinfo {author} {\bibfnamefont {N.~L.}\ \bibnamefont {Nair}}, \bibinfo {author} {\bibfnamefont {W.}~\bibnamefont {Chen}}, \bibinfo {author} {\bibfnamefont {D.}~\bibnamefont {Rees}}, \bibinfo {author} {\bibfnamefont {J.~W.}\ \bibnamefont {Venderbos}}, \bibinfo {author} {\bibfnamefont {R.~M.}\ \bibnamefont {Fernandes}}, \emph {et~al.},\ }\bibfield  {title} {\bibinfo {title} {Three-state nematicity in the triangular lattice antiferromagnet \ce{Fe1/3NbS2}},\ }\href@noop {} {\bibfield  {journal} {\bibinfo  {journal} {Nat. Mater.}\ }\textbf {\bibinfo {volume} {19}},\ \bibinfo {pages} {1062} (\bibinfo {year} {2020})}\BibitemShut {NoStop}%
\bibitem [{\citenamefont {Hwangbo}\ \emph {et~al.}(2024)\citenamefont {Hwangbo}, \citenamefont {Rosenberg}, \citenamefont {Cenker}, \citenamefont {Jiang}, \citenamefont {Wen}, \citenamefont {Xiao}, \citenamefont {Chu},\ and\ \citenamefont {Xu}}]{hwangbo2024strain}%
  \BibitemOpen
  \bibfield  {author} {\bibinfo {author} {\bibfnamefont {K.}~\bibnamefont {Hwangbo}}, \bibinfo {author} {\bibfnamefont {E.}~\bibnamefont {Rosenberg}}, \bibinfo {author} {\bibfnamefont {J.}~\bibnamefont {Cenker}}, \bibinfo {author} {\bibfnamefont {Q.}~\bibnamefont {Jiang}}, \bibinfo {author} {\bibfnamefont {H.}~\bibnamefont {Wen}}, \bibinfo {author} {\bibfnamefont {D.}~\bibnamefont {Xiao}}, \bibinfo {author} {\bibfnamefont {J.-H.}\ \bibnamefont {Chu}},\ and\ \bibinfo {author} {\bibfnamefont {X.}~\bibnamefont {Xu}},\ }\bibfield  {title} {\bibinfo {title} {Strain tuning of vestigial three-state potts nematicity in a correlated antiferromagnet},\ }\href@noop {} {\bibfield  {journal} {\bibinfo  {journal} {Nat. Phys.}\ ,\ \bibinfo {pages} {1}} (\bibinfo {year} {2024})}\BibitemShut {NoStop}%
\bibitem [{\citenamefont {Parkin}\ and\ \citenamefont {Friend}(1980{\natexlab{a}})}]{parkin19803A}%
  \BibitemOpen
  \bibfield  {author} {\bibinfo {author} {\bibfnamefont {S.}~\bibnamefont {Parkin}}\ and\ \bibinfo {author} {\bibfnamefont {R.}~\bibnamefont {Friend}},\ }\bibfield  {title} {\bibinfo {title} {$3d$ transition-metal intercalates of the niobium and tantalum dichalcogenides. {I.} magnetic properties},\ }\href@noop {} {\bibfield  {journal} {\bibinfo  {journal} {Philos. Mag. B}\ }\textbf {\bibinfo {volume} {41}},\ \bibinfo {pages} {65} (\bibinfo {year} {1980}{\natexlab{a}})}\BibitemShut {NoStop}%
\bibitem [{\citenamefont {Parkin}\ and\ \citenamefont {Friend}(1980{\natexlab{b}})}]{parkin19803B}%
  \BibitemOpen
  \bibfield  {author} {\bibinfo {author} {\bibfnamefont {S.}~\bibnamefont {Parkin}}\ and\ \bibinfo {author} {\bibfnamefont {R.}~\bibnamefont {Friend}},\ }\bibfield  {title} {\bibinfo {title} {$3d$ transition-metal intercalates of the niobium and tantalum dichalcogenides. {II.} transport properties},\ }\href@noop {} {\bibfield  {journal} {\bibinfo  {journal} {Philos. Mag. B}\ }\textbf {\bibinfo {volume} {41}},\ \bibinfo {pages} {95} (\bibinfo {year} {1980}{\natexlab{b}})}\BibitemShut {NoStop}%
\bibitem [{\citenamefont {Miyadai}\ \emph {et~al.}(1983)\citenamefont {Miyadai}, \citenamefont {Kikuchi}, \citenamefont {Kondo}, \citenamefont {Sakka}, \citenamefont {Arai},\ and\ \citenamefont {Ishikawa}}]{miyadai1983magnetic}%
  \BibitemOpen
  \bibfield  {author} {\bibinfo {author} {\bibfnamefont {T.}~\bibnamefont {Miyadai}}, \bibinfo {author} {\bibfnamefont {K.}~\bibnamefont {Kikuchi}}, \bibinfo {author} {\bibfnamefont {H.}~\bibnamefont {Kondo}}, \bibinfo {author} {\bibfnamefont {S.}~\bibnamefont {Sakka}}, \bibinfo {author} {\bibfnamefont {M.}~\bibnamefont {Arai}},\ and\ \bibinfo {author} {\bibfnamefont {Y.}~\bibnamefont {Ishikawa}},\ }\bibfield  {title} {\bibinfo {title} {Magnetic properties of \ce{Cr1/3NbS2}},\ }\href@noop {} {\bibfield  {journal} {\bibinfo  {journal} {J. Phys. Soc. Jpn.}\ }\textbf {\bibinfo {volume} {52}},\ \bibinfo {pages} {1394} (\bibinfo {year} {1983})}\BibitemShut {NoStop}%
\bibitem [{\citenamefont {Morosan}\ \emph {et~al.}(2007)\citenamefont {Morosan}, \citenamefont {Zandbergen}, \citenamefont {Li}, \citenamefont {Lee}, \citenamefont {Checkelsky}, \citenamefont {Heinrich}, \citenamefont {Siegrist}, \citenamefont {Ong},\ and\ \citenamefont {Cava}}]{morosan2007sharp}%
  \BibitemOpen
  \bibfield  {author} {\bibinfo {author} {\bibfnamefont {E.}~\bibnamefont {Morosan}}, \bibinfo {author} {\bibfnamefont {H.}~\bibnamefont {Zandbergen}}, \bibinfo {author} {\bibfnamefont {L.}~\bibnamefont {Li}}, \bibinfo {author} {\bibfnamefont {M.}~\bibnamefont {Lee}}, \bibinfo {author} {\bibfnamefont {J.}~\bibnamefont {Checkelsky}}, \bibinfo {author} {\bibfnamefont {M.}~\bibnamefont {Heinrich}}, \bibinfo {author} {\bibfnamefont {T.}~\bibnamefont {Siegrist}}, \bibinfo {author} {\bibfnamefont {N.~P.}\ \bibnamefont {Ong}},\ and\ \bibinfo {author} {\bibfnamefont {R.}~\bibnamefont {Cava}},\ }\bibfield  {title} {\bibinfo {title} {Sharp switching of the magnetization in \ce{Fe1/4TaS2}},\ }\href@noop {} {\bibfield  {journal} {\bibinfo  {journal} {Phys. Rev. B}\ }\textbf {\bibinfo {volume} {75}},\ \bibinfo {pages} {104401} (\bibinfo {year} {2007})}\BibitemShut {NoStop}%
\bibitem [{\citenamefont {Wu}\ \emph {et~al.}(2022)\citenamefont {Wu}, \citenamefont {Xu}, \citenamefont {Haley}, \citenamefont {Weber}, \citenamefont {Acharya}, \citenamefont {Maniv}, \citenamefont {Qiu}, \citenamefont {Aczel}, \citenamefont {Settineri}, \citenamefont {Neaton} \emph {et~al.}}]{wu2022highly}%
  \BibitemOpen
  \bibfield  {author} {\bibinfo {author} {\bibfnamefont {S.}~\bibnamefont {Wu}}, \bibinfo {author} {\bibfnamefont {Z.}~\bibnamefont {Xu}}, \bibinfo {author} {\bibfnamefont {S.~C.}\ \bibnamefont {Haley}}, \bibinfo {author} {\bibfnamefont {S.~F.}\ \bibnamefont {Weber}}, \bibinfo {author} {\bibfnamefont {A.}~\bibnamefont {Acharya}}, \bibinfo {author} {\bibfnamefont {E.}~\bibnamefont {Maniv}}, \bibinfo {author} {\bibfnamefont {Y.}~\bibnamefont {Qiu}}, \bibinfo {author} {\bibfnamefont {A.~A.}\ \bibnamefont {Aczel}}, \bibinfo {author} {\bibfnamefont {N.~S.}\ \bibnamefont {Settineri}}, \bibinfo {author} {\bibfnamefont {J.~B.}\ \bibnamefont {Neaton}}, \emph {et~al.},\ }\bibfield  {title} {\bibinfo {title} {Highly tunable magnetic phases in transition-metal dichalcogenide \ce{Fe_{1/3+$\delta$}NbS2}},\ }\href@noop {} {\bibfield  {journal} {\bibinfo  {journal} {Phys. Rev. X}\ }\textbf {\bibinfo {volume} {12}},\ \bibinfo {pages} {021003} (\bibinfo {year} {2022})}\BibitemShut {NoStop}%
\bibitem [{\citenamefont {Xie}\ \emph {et~al.}(2022)\citenamefont {Xie}, \citenamefont {Husremovic}, \citenamefont {Gonzalez}, \citenamefont {Craig},\ and\ \citenamefont {Bediako}}]{xie2022structure}%
  \BibitemOpen
  \bibfield  {author} {\bibinfo {author} {\bibfnamefont {L.~S.}\ \bibnamefont {Xie}}, \bibinfo {author} {\bibfnamefont {S.}~\bibnamefont {Husremovic}}, \bibinfo {author} {\bibfnamefont {O.}~\bibnamefont {Gonzalez}}, \bibinfo {author} {\bibfnamefont {I.~M.}\ \bibnamefont {Craig}},\ and\ \bibinfo {author} {\bibfnamefont {D.~K.}\ \bibnamefont {Bediako}},\ }\bibfield  {title} {\bibinfo {title} {Structure and magnetism of iron-and chromium-intercalated niobium and tantalum disulfides},\ }\href@noop {} {\bibfield  {journal} {\bibinfo  {journal} {J. Am. Chem. Soc.}\ }\textbf {\bibinfo {volume} {144}},\ \bibinfo {pages} {9525} (\bibinfo {year} {2022})}\BibitemShut {NoStop}%
\bibitem [{\citenamefont {Park}\ \emph {et~al.}(2024{\natexlab{a}})\citenamefont {Park}, \citenamefont {Cho}, \citenamefont {Kim}, \citenamefont {An}, \citenamefont {Avdeev}, \citenamefont {Iida}, \citenamefont {Kajimoto},\ and\ \citenamefont {Park}}]{park2024composition}%
  \BibitemOpen
  \bibfield  {author} {\bibinfo {author} {\bibfnamefont {P.}~\bibnamefont {Park}}, \bibinfo {author} {\bibfnamefont {W.}~\bibnamefont {Cho}}, \bibinfo {author} {\bibfnamefont {C.}~\bibnamefont {Kim}}, \bibinfo {author} {\bibfnamefont {Y.}~\bibnamefont {An}}, \bibinfo {author} {\bibfnamefont {M.}~\bibnamefont {Avdeev}}, \bibinfo {author} {\bibfnamefont {K.}~\bibnamefont {Iida}}, \bibinfo {author} {\bibfnamefont {R.}~\bibnamefont {Kajimoto}},\ and\ \bibinfo {author} {\bibfnamefont {J.-G.}\ \bibnamefont {Park}},\ }\bibfield  {title} {\bibinfo {title} {Composition dependence of bulk properties in the {C}o-intercalated transition metal dichalcogenide \ce{Co1/3TaS2}},\ }\href@noop {} {\bibfield  {journal} {\bibinfo  {journal} {Phys. Rev. B}\ }\textbf {\bibinfo {volume} {109}},\ \bibinfo {pages} {L060403} (\bibinfo {year} {2024}{\natexlab{a}})}\BibitemShut {NoStop}%
\bibitem [{\citenamefont {{van den Berg}}\ and\ \citenamefont {Cossee}(1968)}]{VANDENBERG1968143}%
  \BibitemOpen
  \bibfield  {author} {\bibinfo {author} {\bibfnamefont {J.}~\bibnamefont {{van den Berg}}}\ and\ \bibinfo {author} {\bibfnamefont {P.}~\bibnamefont {Cossee}},\ }\bibfield  {title} {\bibinfo {title} {Structural aspects and magnetic behaviour of \ce{NbS2} and \ce{TaS2} containing extra metal atoms of the first transition series},\ }\href@noop {} {\bibfield  {journal} {\bibinfo  {journal} {Inorganica Chimica Acta}\ }\textbf {\bibinfo {volume} {2}},\ \bibinfo {pages} {143} (\bibinfo {year} {1968})}\BibitemShut {NoStop}%
\bibitem [{\citenamefont {Ghimire}\ \emph {et~al.}(2018)\citenamefont {Ghimire}, \citenamefont {Botana}, \citenamefont {Jiang}, \citenamefont {Zhang}, \citenamefont {Chen},\ and\ \citenamefont {Mitchell}}]{Ghimire2018-bn}%
  \BibitemOpen
  \bibfield  {author} {\bibinfo {author} {\bibfnamefont {N.~J.}\ \bibnamefont {Ghimire}}, \bibinfo {author} {\bibfnamefont {A.~S.}\ \bibnamefont {Botana}}, \bibinfo {author} {\bibfnamefont {J.~S.}\ \bibnamefont {Jiang}}, \bibinfo {author} {\bibfnamefont {J.}~\bibnamefont {Zhang}}, \bibinfo {author} {\bibfnamefont {Y.-S.}\ \bibnamefont {Chen}},\ and\ \bibinfo {author} {\bibfnamefont {J.~F.}\ \bibnamefont {Mitchell}},\ }\bibfield  {title} {\bibinfo {title} {Large anomalous hall effect in the chiral-lattice antiferromagnet \ce{CoNb3S6}},\ }\href@noop {} {\bibfield  {journal} {\bibinfo  {journal} {Nat. Commun.}\ }\textbf {\bibinfo {volume} {9}},\ \bibinfo {pages} {3280} (\bibinfo {year} {2018})}\BibitemShut {NoStop}%
\bibitem [{\citenamefont {Park}\ \emph {et~al.}(2022)\citenamefont {Park}, \citenamefont {Kang}, \citenamefont {Kim}, \citenamefont {Lee}, \citenamefont {Noh}, \citenamefont {Han},\ and\ \citenamefont {Park}}]{park2022field}%
  \BibitemOpen
  \bibfield  {author} {\bibinfo {author} {\bibfnamefont {P.}~\bibnamefont {Park}}, \bibinfo {author} {\bibfnamefont {Y.-G.}\ \bibnamefont {Kang}}, \bibinfo {author} {\bibfnamefont {J.}~\bibnamefont {Kim}}, \bibinfo {author} {\bibfnamefont {K.~H.}\ \bibnamefont {Lee}}, \bibinfo {author} {\bibfnamefont {H.-J.}\ \bibnamefont {Noh}}, \bibinfo {author} {\bibfnamefont {M.~J.}\ \bibnamefont {Han}},\ and\ \bibinfo {author} {\bibfnamefont {J.-G.}\ \bibnamefont {Park}},\ }\bibfield  {title} {\bibinfo {title} {Field-tunable toroidal moment and anomalous hall effect in noncollinear antiferromagnetic weyl semimetal \ce{Co1/3TaS2}},\ }\href@noop {} {\bibfield  {journal} {\bibinfo  {journal} {npj Quantum Materials}\ }\textbf {\bibinfo {volume} {7}},\ \bibinfo {pages} {42} (\bibinfo {year} {2022})}\BibitemShut {NoStop}%
\bibitem [{\citenamefont {Mangelsen}\ \emph {et~al.}(2021)\citenamefont {Mangelsen}, \citenamefont {Zimmer}, \citenamefont {N\"ather}, \citenamefont {Mankovsky}, \citenamefont {Polesya}, \citenamefont {Ebert},\ and\ \citenamefont {Bensch}}]{PhysRevB.103.184408}%
  \BibitemOpen
  \bibfield  {author} {\bibinfo {author} {\bibfnamefont {S.}~\bibnamefont {Mangelsen}}, \bibinfo {author} {\bibfnamefont {P.}~\bibnamefont {Zimmer}}, \bibinfo {author} {\bibfnamefont {C.}~\bibnamefont {N\"ather}}, \bibinfo {author} {\bibfnamefont {S.}~\bibnamefont {Mankovsky}}, \bibinfo {author} {\bibfnamefont {S.}~\bibnamefont {Polesya}}, \bibinfo {author} {\bibfnamefont {H.}~\bibnamefont {Ebert}},\ and\ \bibinfo {author} {\bibfnamefont {W.}~\bibnamefont {Bensch}},\ }\bibfield  {title} {\bibinfo {title} {Interplay of sample composition and anomalous hall effect in \ce{Co_xNbS2}},\ }\href@noop {} {\bibfield  {journal} {\bibinfo  {journal} {Phys. Rev. B}\ }\textbf {\bibinfo {volume} {103}},\ \bibinfo {pages} {184408} (\bibinfo {year} {2021})}\BibitemShut {NoStop}%
\bibitem [{\citenamefont {Tenasini}\ \emph {et~al.}(2020)\citenamefont {Tenasini}, \citenamefont {Martino}, \citenamefont {Ubrig}, \citenamefont {Ghimire}, \citenamefont {Berger}, \citenamefont {Zaharko}, \citenamefont {Wu}, \citenamefont {Mitchell}, \citenamefont {Martin}, \citenamefont {Forr\'o},\ and\ \citenamefont {Morpurgo}}]{PhysRevResearch.2.023051}%
  \BibitemOpen
  \bibfield  {author} {\bibinfo {author} {\bibfnamefont {G.}~\bibnamefont {Tenasini}}, \bibinfo {author} {\bibfnamefont {E.}~\bibnamefont {Martino}}, \bibinfo {author} {\bibfnamefont {N.}~\bibnamefont {Ubrig}}, \bibinfo {author} {\bibfnamefont {N.~J.}\ \bibnamefont {Ghimire}}, \bibinfo {author} {\bibfnamefont {H.}~\bibnamefont {Berger}}, \bibinfo {author} {\bibfnamefont {O.}~\bibnamefont {Zaharko}}, \bibinfo {author} {\bibfnamefont {F.}~\bibnamefont {Wu}}, \bibinfo {author} {\bibfnamefont {J.~F.}\ \bibnamefont {Mitchell}}, \bibinfo {author} {\bibfnamefont {I.}~\bibnamefont {Martin}}, \bibinfo {author} {\bibfnamefont {L.}~\bibnamefont {Forr\'o}},\ and\ \bibinfo {author} {\bibfnamefont {A.~F.}\ \bibnamefont {Morpurgo}},\ }\bibfield  {title} {\bibinfo {title} {Giant anomalous hall effect in quasi-two-dimensional layered antiferromagnet \ce{Co_{1/3}NbS2}},\ }\href@noop {} {\bibfield  {journal} {\bibinfo  {journal} {Phys. Rev. Res.}\ }\textbf {\bibinfo {volume} {2}},\ \bibinfo {pages} {023051} (\bibinfo {year}
  {2020})}\BibitemShut {NoStop}%
\bibitem [{\citenamefont {Tanaka}\ \emph {et~al.}(2022)\citenamefont {Tanaka}, \citenamefont {Okazaki}, \citenamefont {Kuroda}, \citenamefont {Noguchi}, \citenamefont {Arai}, \citenamefont {Minami}, \citenamefont {Ideta}, \citenamefont {Tanaka}, \citenamefont {Lu}, \citenamefont {Hashimoto} \emph {et~al.}}]{tanaka2022large}%
  \BibitemOpen
  \bibfield  {author} {\bibinfo {author} {\bibfnamefont {H.}~\bibnamefont {Tanaka}}, \bibinfo {author} {\bibfnamefont {S.}~\bibnamefont {Okazaki}}, \bibinfo {author} {\bibfnamefont {K.}~\bibnamefont {Kuroda}}, \bibinfo {author} {\bibfnamefont {R.}~\bibnamefont {Noguchi}}, \bibinfo {author} {\bibfnamefont {Y.}~\bibnamefont {Arai}}, \bibinfo {author} {\bibfnamefont {S.}~\bibnamefont {Minami}}, \bibinfo {author} {\bibfnamefont {S.}~\bibnamefont {Ideta}}, \bibinfo {author} {\bibfnamefont {K.}~\bibnamefont {Tanaka}}, \bibinfo {author} {\bibfnamefont {D.}~\bibnamefont {Lu}}, \bibinfo {author} {\bibfnamefont {M.}~\bibnamefont {Hashimoto}}, \emph {et~al.},\ }\bibfield  {title} {\bibinfo {title} {Large anomalous hall effect induced by weak ferromagnetism in the noncentrosymmetric antiferromagnet \ce{CoNb3S6}},\ }\href@noop {} {\bibfield  {journal} {\bibinfo  {journal} {Phys. Rev. B}\ }\textbf {\bibinfo {volume} {105}},\ \bibinfo {pages} {L121102} (\bibinfo {year} {2022})}\BibitemShut {NoStop}%
\bibitem [{\citenamefont {Park}\ \emph {et~al.}(2023)\citenamefont {Park}, \citenamefont {Cho}, \citenamefont {Kim}, \citenamefont {An}, \citenamefont {Kang}, \citenamefont {Avdeev}, \citenamefont {Sibille}, \citenamefont {Iida}, \citenamefont {Kajimoto}, \citenamefont {Lee} \emph {et~al.}}]{park2023tetrahedral}%
  \BibitemOpen
  \bibfield  {author} {\bibinfo {author} {\bibfnamefont {P.}~\bibnamefont {Park}}, \bibinfo {author} {\bibfnamefont {W.}~\bibnamefont {Cho}}, \bibinfo {author} {\bibfnamefont {C.}~\bibnamefont {Kim}}, \bibinfo {author} {\bibfnamefont {Y.}~\bibnamefont {An}}, \bibinfo {author} {\bibfnamefont {Y.-G.}\ \bibnamefont {Kang}}, \bibinfo {author} {\bibfnamefont {M.}~\bibnamefont {Avdeev}}, \bibinfo {author} {\bibfnamefont {R.}~\bibnamefont {Sibille}}, \bibinfo {author} {\bibfnamefont {K.}~\bibnamefont {Iida}}, \bibinfo {author} {\bibfnamefont {R.}~\bibnamefont {Kajimoto}}, \bibinfo {author} {\bibfnamefont {K.~H.}\ \bibnamefont {Lee}}, \emph {et~al.},\ }\bibfield  {title} {\bibinfo {title} {Tetrahedral triple-q magnetic ordering and large spontaneous hall conductivity in the metallic triangular antiferromagnet \ce{Co1/3TaS2}},\ }\href@noop {} {\bibfield  {journal} {\bibinfo  {journal} {Nat. Commun.}\ }\textbf {\bibinfo {volume} {14}},\ \bibinfo {pages} {8346} (\bibinfo {year} {2023})}\BibitemShut {NoStop}%
\bibitem [{\citenamefont {Takagi}\ \emph {et~al.}(2023)\citenamefont {Takagi}, \citenamefont {Takagi}, \citenamefont {Minami}, \citenamefont {Nomoto}, \citenamefont {Ohishi}, \citenamefont {Suzuki}, \citenamefont {Yanagi}, \citenamefont {Hirayama}, \citenamefont {Khanh}, \citenamefont {Karube} \emph {et~al.}}]{takagi2023spontaneous}%
  \BibitemOpen
  \bibfield  {author} {\bibinfo {author} {\bibfnamefont {H.}~\bibnamefont {Takagi}}, \bibinfo {author} {\bibfnamefont {R.}~\bibnamefont {Takagi}}, \bibinfo {author} {\bibfnamefont {S.}~\bibnamefont {Minami}}, \bibinfo {author} {\bibfnamefont {T.}~\bibnamefont {Nomoto}}, \bibinfo {author} {\bibfnamefont {K.}~\bibnamefont {Ohishi}}, \bibinfo {author} {\bibfnamefont {M.-T.}\ \bibnamefont {Suzuki}}, \bibinfo {author} {\bibfnamefont {Y.}~\bibnamefont {Yanagi}}, \bibinfo {author} {\bibfnamefont {M.}~\bibnamefont {Hirayama}}, \bibinfo {author} {\bibfnamefont {N.}~\bibnamefont {Khanh}}, \bibinfo {author} {\bibfnamefont {K.}~\bibnamefont {Karube}}, \emph {et~al.},\ }\bibfield  {title} {\bibinfo {title} {Spontaneous topological hall effect induced by non-coplanar antiferromagnetic order in intercalated van der waals materials},\ }\href@noop {} {\bibfield  {journal} {\bibinfo  {journal} {Nat. Phys.}\ }\textbf {\bibinfo {volume} {19}},\ \bibinfo {pages} {961} (\bibinfo {year} {2023})}\BibitemShut {NoStop}%
\bibitem [{\citenamefont {Park}\ \emph {et~al.}(2024{\natexlab{b}})\citenamefont {Park}, \citenamefont {Cho}, \citenamefont {Kim}, \citenamefont {An}, \citenamefont {Iida}, \citenamefont {Kajimoto}, \citenamefont {Matin}, \citenamefont {Zhang}, \citenamefont {Batista},\ and\ \citenamefont {Park}}]{Park2024-nl}%
  \BibitemOpen
  \bibfield  {author} {\bibinfo {author} {\bibfnamefont {P.}~\bibnamefont {Park}}, \bibinfo {author} {\bibfnamefont {W.}~\bibnamefont {Cho}}, \bibinfo {author} {\bibfnamefont {C.}~\bibnamefont {Kim}}, \bibinfo {author} {\bibfnamefont {Y.}~\bibnamefont {An}}, \bibinfo {author} {\bibfnamefont {K.}~\bibnamefont {Iida}}, \bibinfo {author} {\bibfnamefont {R.}~\bibnamefont {Kajimoto}}, \bibinfo {author} {\bibfnamefont {S.}~\bibnamefont {Matin}}, \bibinfo {author} {\bibfnamefont {S.-S.}\ \bibnamefont {Zhang}}, \bibinfo {author} {\bibfnamefont {C.~D.}\ \bibnamefont {Batista}},\ and\ \bibinfo {author} {\bibfnamefont {J.-G.}\ \bibnamefont {Park}},\ }\bibfield  {title} {\bibinfo {title} {Contrasting dynamical properties of single-{Q} and triple-{Q} magnetic orderings in a triangular lattice antiferromagnet},\ }\href@noop {} {\bibfield  {journal} {\bibinfo  {journal} {arXiv [cond-mat.str-el]}\ } (\bibinfo {year} {2024}{\natexlab{b}})}\BibitemShut {NoStop}%
\bibitem [{\citenamefont {Yanagi}\ \emph {et~al.}(2023)\citenamefont {Yanagi}, \citenamefont {Kusunose}, \citenamefont {Nomoto}, \citenamefont {Arita},\ and\ \citenamefont {Suzuki}}]{yanagi2023generation}%
  \BibitemOpen
  \bibfield  {author} {\bibinfo {author} {\bibfnamefont {Y.}~\bibnamefont {Yanagi}}, \bibinfo {author} {\bibfnamefont {H.}~\bibnamefont {Kusunose}}, \bibinfo {author} {\bibfnamefont {T.}~\bibnamefont {Nomoto}}, \bibinfo {author} {\bibfnamefont {R.}~\bibnamefont {Arita}},\ and\ \bibinfo {author} {\bibfnamefont {M.-T.}\ \bibnamefont {Suzuki}},\ }\bibfield  {title} {\bibinfo {title} {Generation of modulated magnetic structures based on cluster multipole expansion: Application to $\alpha$-\ce{Mn} and \ce{CoM3S6}},\ }\href@noop {} {\bibfield  {journal} {\bibinfo  {journal} {Phys. Rev. B}\ }\textbf {\bibinfo {volume} {107}},\ \bibinfo {pages} {014407} (\bibinfo {year} {2023})}\BibitemShut {NoStop}%
\bibitem [{\citenamefont {Park}\ and\ \citenamefont {Martin}(2024)}]{park2024dft+}%
  \BibitemOpen
  \bibfield  {author} {\bibinfo {author} {\bibfnamefont {H.}~\bibnamefont {Park}}\ and\ \bibinfo {author} {\bibfnamefont {I.}~\bibnamefont {Martin}},\ }\bibfield  {title} {\bibinfo {title} {{DFT+ DMFT} study of the magnetic susceptibility and the correlated electronic structure in transition-metal intercalated \ce{NbS2}},\ }\href@noop {} {\bibfield  {journal} {\bibinfo  {journal} {Phys. Rev. B}\ }\textbf {\bibinfo {volume} {109}},\ \bibinfo {pages} {085110} (\bibinfo {year} {2024})}\BibitemShut {NoStop}%
\bibitem [{\citenamefont {Heinonen}\ \emph {et~al.}(2022)\citenamefont {Heinonen}, \citenamefont {Heinonen},\ and\ \citenamefont {Park}}]{heinonen2022magnetic}%
  \BibitemOpen
  \bibfield  {author} {\bibinfo {author} {\bibfnamefont {O.}~\bibnamefont {Heinonen}}, \bibinfo {author} {\bibfnamefont {R.}~\bibnamefont {Heinonen}},\ and\ \bibinfo {author} {\bibfnamefont {H.}~\bibnamefont {Park}},\ }\bibfield  {title} {\bibinfo {title} {Magnetic ground states of a model for \ce{MNb3S6} (\ce{ M= Co, Fe, Ni})},\ }\href@noop {} {\bibfield  {journal} {\bibinfo  {journal} {Phys. Rev. Materials}\ }\textbf {\bibinfo {volume} {6}},\ \bibinfo {pages} {024405} (\bibinfo {year} {2022})}\BibitemShut {NoStop}%
\bibitem [{\citenamefont {Chu}\ \emph {et~al.}(2012)\citenamefont {Chu}, \citenamefont {Kuo}, \citenamefont {Analytis},\ and\ \citenamefont {Fisher}}]{Chu2012-vj}%
  \BibitemOpen
  \bibfield  {author} {\bibinfo {author} {\bibfnamefont {J.-H.}\ \bibnamefont {Chu}}, \bibinfo {author} {\bibfnamefont {H.-H.}\ \bibnamefont {Kuo}}, \bibinfo {author} {\bibfnamefont {J.~G.}\ \bibnamefont {Analytis}},\ and\ \bibinfo {author} {\bibfnamefont {I.~R.}\ \bibnamefont {Fisher}},\ }\bibfield  {title} {\bibinfo {title} {Divergent nematic susceptibility in an iron arsenide superconductor},\ }\href@noop {} {\bibfield  {journal} {\bibinfo  {journal} {Science}\ }\textbf {\bibinfo {volume} {337}},\ \bibinfo {pages} {710} (\bibinfo {year} {2012})}\BibitemShut {NoStop}%
\bibitem [{\citenamefont {Amin}(1991)}]{amin1991elastoresistance}%
  \BibitemOpen
  \bibfield  {author} {\bibinfo {author} {\bibfnamefont {A.}~\bibnamefont {Amin}},\ }\bibfield  {title} {\bibinfo {title} {Elastoresistance tensor components for thick-film resistors},\ }\href@noop {} {\bibfield  {journal} {\bibinfo  {journal} {Applied physics Lett.}\ }\textbf {\bibinfo {volume} {58}},\ \bibinfo {pages} {1446} (\bibinfo {year} {1991})}\BibitemShut {NoStop}%
\bibitem [{\citenamefont {Mutch}\ \emph {et~al.}(2019)\citenamefont {Mutch}, \citenamefont {Chen}, \citenamefont {Went}, \citenamefont {Qian}, \citenamefont {Wilson}, \citenamefont {Andreev}, \citenamefont {Chen},\ and\ \citenamefont {Chu}}]{mutch2019evidence}%
  \BibitemOpen
  \bibfield  {author} {\bibinfo {author} {\bibfnamefont {J.}~\bibnamefont {Mutch}}, \bibinfo {author} {\bibfnamefont {W.-C.}\ \bibnamefont {Chen}}, \bibinfo {author} {\bibfnamefont {P.}~\bibnamefont {Went}}, \bibinfo {author} {\bibfnamefont {T.}~\bibnamefont {Qian}}, \bibinfo {author} {\bibfnamefont {I.~Z.}\ \bibnamefont {Wilson}}, \bibinfo {author} {\bibfnamefont {A.}~\bibnamefont {Andreev}}, \bibinfo {author} {\bibfnamefont {C.-C.}\ \bibnamefont {Chen}},\ and\ \bibinfo {author} {\bibfnamefont {J.-H.}\ \bibnamefont {Chu}},\ }\bibfield  {title} {\bibinfo {title} {Evidence for a strain-tuned topological phase transition in \ce{ZrTe5}},\ }\href@noop {} {\bibfield  {journal} {\bibinfo  {journal} {Sci. Adv.}\ }\textbf {\bibinfo {volume} {5}},\ \bibinfo {pages} {eaav9771} (\bibinfo {year} {2019})}\BibitemShut {NoStop}%
\bibitem [{\citenamefont {Bartlett}\ \emph {et~al.}(2021)\citenamefont {Bartlett}, \citenamefont {Steppke}, \citenamefont {Hosoi}, \citenamefont {Noad}, \citenamefont {Park}, \citenamefont {Timm}, \citenamefont {Shibauchi}, \citenamefont {Mackenzie},\ and\ \citenamefont {Hicks}}]{Bartlett2021-hv}%
  \BibitemOpen
  \bibfield  {author} {\bibinfo {author} {\bibfnamefont {J.~M.}\ \bibnamefont {Bartlett}}, \bibinfo {author} {\bibfnamefont {A.}~\bibnamefont {Steppke}}, \bibinfo {author} {\bibfnamefont {S.}~\bibnamefont {Hosoi}}, \bibinfo {author} {\bibfnamefont {H.}~\bibnamefont {Noad}}, \bibinfo {author} {\bibfnamefont {J.}~\bibnamefont {Park}}, \bibinfo {author} {\bibfnamefont {C.}~\bibnamefont {Timm}}, \bibinfo {author} {\bibfnamefont {T.}~\bibnamefont {Shibauchi}}, \bibinfo {author} {\bibfnamefont {A.~P.}\ \bibnamefont {Mackenzie}},\ and\ \bibinfo {author} {\bibfnamefont {C.~W.}\ \bibnamefont {Hicks}},\ }\bibfield  {title} {\bibinfo {title} {Relationship between transport anisotropy and nematicity in {FeSe}},\ }\href@noop {} {\bibfield  {journal} {\bibinfo  {journal} {Phys. Rev. X}\ }\textbf {\bibinfo {volume} {11}},\ \bibinfo {pages} {021038} (\bibinfo {year} {2021})}\BibitemShut {NoStop}%
\bibitem [{\citenamefont {Lilly}\ \emph {et~al.}(1999{\natexlab{b}})\citenamefont {Lilly}, \citenamefont {Cooper}, \citenamefont {Eisenstein}, \citenamefont {Pfeiffer},\ and\ \citenamefont {West}}]{Lilly1999-ix}%
  \BibitemOpen
  \bibfield  {author} {\bibinfo {author} {\bibfnamefont {M.~P.}\ \bibnamefont {Lilly}}, \bibinfo {author} {\bibfnamefont {K.~B.}\ \bibnamefont {Cooper}}, \bibinfo {author} {\bibfnamefont {J.~P.}\ \bibnamefont {Eisenstein}}, \bibinfo {author} {\bibfnamefont {L.~N.}\ \bibnamefont {Pfeiffer}},\ and\ \bibinfo {author} {\bibfnamefont {K.~W.}\ \bibnamefont {West}},\ }\bibfield  {title} {\bibinfo {title} {Anisotropic states of two-dimensional electron systems in high landau levels: Effect of an in-plane magnetic field},\ }\href@noop {} {\bibfield  {journal} {\bibinfo  {journal} {Phys. Rev. Lett.}\ }\textbf {\bibinfo {volume} {83}},\ \bibinfo {pages} {824} (\bibinfo {year} {1999}{\natexlab{b}})}\BibitemShut {NoStop}%
\bibitem [{\citenamefont {Ye}\ and\ \citenamefont {Chubukov}(2017)}]{Ye2017-ax}%
  \BibitemOpen
  \bibfield  {author} {\bibinfo {author} {\bibfnamefont {M.}~\bibnamefont {Ye}}\ and\ \bibinfo {author} {\bibfnamefont {A.~V.}\ \bibnamefont {Chubukov}},\ }\bibfield  {title} {\bibinfo {title} {Half-magnetization plateau in a heisenberg antiferromagnet on a triangular lattice},\ }\href@noop {} {\bibfield  {journal} {\bibinfo  {journal} {Phys. Rev. B}\ }\textbf {\bibinfo {volume} {96}},\ \bibinfo {pages} {140406} (\bibinfo {year} {2017})}\BibitemShut {NoStop}%
\bibitem [{\citenamefont {Fernandes}\ \emph {et~al.}(2014)\citenamefont {Fernandes}, \citenamefont {Chubukov},\ and\ \citenamefont {Schmalian}}]{Fernandes2014-qg}%
  \BibitemOpen
  \bibfield  {author} {\bibinfo {author} {\bibfnamefont {R.~M.}\ \bibnamefont {Fernandes}}, \bibinfo {author} {\bibfnamefont {A.~V.}\ \bibnamefont {Chubukov}},\ and\ \bibinfo {author} {\bibfnamefont {J.}~\bibnamefont {Schmalian}},\ }\bibfield  {title} {\bibinfo {title} {What drives nematic order in iron-based superconductors?},\ }\href@noop {} {\bibfield  {journal} {\bibinfo  {journal} {Nat. Phys.}\ }\textbf {\bibinfo {volume} {10}},\ \bibinfo {pages} {97} (\bibinfo {year} {2014})}\BibitemShut {NoStop}%
\bibitem [{\citenamefont {Nie}\ \emph {et~al.}(2014)\citenamefont {Nie}, \citenamefont {Tarjus},\ and\ \citenamefont {Kivelson}}]{Nie2014-sz}%
  \BibitemOpen
  \bibfield  {author} {\bibinfo {author} {\bibfnamefont {L.}~\bibnamefont {Nie}}, \bibinfo {author} {\bibfnamefont {G.}~\bibnamefont {Tarjus}},\ and\ \bibinfo {author} {\bibfnamefont {S.~A.}\ \bibnamefont {Kivelson}},\ }\bibfield  {title} {\bibinfo {title} {Quenched disorder and vestigial nematicity in the pseudogap regime of the cuprates},\ }\href@noop {} {\bibfield  {journal} {\bibinfo  {journal} {Proc. Natl. Acad. Sci. U. S. A.}\ }\textbf {\bibinfo {volume} {111}},\ \bibinfo {pages} {7980} (\bibinfo {year} {2014})}\BibitemShut {NoStop}%
\bibitem [{\citenamefont {Nie}\ \emph {et~al.}(2022)\citenamefont {Nie}, \citenamefont {Sun}, \citenamefont {Ma}, \citenamefont {Song}, \citenamefont {Zheng}, \citenamefont {Liang}, \citenamefont {Wu}, \citenamefont {Yu}, \citenamefont {Li}, \citenamefont {Shan}, \citenamefont {Zhao}, \citenamefont {Li}, \citenamefont {Kang}, \citenamefont {Wu}, \citenamefont {Zhou}, \citenamefont {Liu}, \citenamefont {Xiang}, \citenamefont {Ying}, \citenamefont {Wang}, \citenamefont {Wu},\ and\ \citenamefont {Chen}}]{Nie2022-tx}%
  \BibitemOpen
  \bibfield  {author} {\bibinfo {author} {\bibfnamefont {L.}~\bibnamefont {Nie}}, \bibinfo {author} {\bibfnamefont {K.}~\bibnamefont {Sun}}, \bibinfo {author} {\bibfnamefont {W.}~\bibnamefont {Ma}}, \bibinfo {author} {\bibfnamefont {D.}~\bibnamefont {Song}}, \bibinfo {author} {\bibfnamefont {L.}~\bibnamefont {Zheng}}, \bibinfo {author} {\bibfnamefont {Z.}~\bibnamefont {Liang}}, \bibinfo {author} {\bibfnamefont {P.}~\bibnamefont {Wu}}, \bibinfo {author} {\bibfnamefont {F.}~\bibnamefont {Yu}}, \bibinfo {author} {\bibfnamefont {J.}~\bibnamefont {Li}}, \bibinfo {author} {\bibfnamefont {M.}~\bibnamefont {Shan}}, \bibinfo {author} {\bibfnamefont {D.}~\bibnamefont {Zhao}}, \bibinfo {author} {\bibfnamefont {S.}~\bibnamefont {Li}}, \bibinfo {author} {\bibfnamefont {B.}~\bibnamefont {Kang}}, \bibinfo {author} {\bibfnamefont {Z.}~\bibnamefont {Wu}}, \bibinfo {author} {\bibfnamefont {Y.}~\bibnamefont {Zhou}}, \bibinfo {author} {\bibfnamefont {K.}~\bibnamefont {Liu}}, \bibinfo {author} {\bibfnamefont {Z.}~\bibnamefont
  {Xiang}}, \bibinfo {author} {\bibfnamefont {J.}~\bibnamefont {Ying}}, \bibinfo {author} {\bibfnamefont {Z.}~\bibnamefont {Wang}}, \bibinfo {author} {\bibfnamefont {T.}~\bibnamefont {Wu}},\ and\ \bibinfo {author} {\bibfnamefont {X.}~\bibnamefont {Chen}},\ }\bibfield  {title} {\bibinfo {title} {Charge-density-wave-driven electronic nematicity in a kagome superconductor},\ }\href@noop {} {\bibfield  {journal} {\bibinfo  {journal} {Nature}\ }\textbf {\bibinfo {volume} {604}},\ \bibinfo {pages} {59} (\bibinfo {year} {2022})}\BibitemShut {NoStop}%
\bibitem [{\citenamefont {Kar}\ \emph {et~al.}(2025)\citenamefont {Kar}, \citenamefont {Basak}, \citenamefont {Li}, \citenamefont {Korshunov}, \citenamefont {Subires}, \citenamefont {Phillips}, \citenamefont {Lim}, \citenamefont {Zhou}, \citenamefont {Song}, \citenamefont {Wang}, \citenamefont {Lau}, \citenamefont {Garbarino}, \citenamefont {Gargiani}, \citenamefont {Zhao}, \citenamefont {Plueckthun}, \citenamefont {Francoual}, \citenamefont {Jana}, \citenamefont {Vobornik}, \citenamefont {Valla}, \citenamefont {Rajapitamahuni}, \citenamefont {Analytis}, \citenamefont {Birgeneau}, \citenamefont {Vescovo}, \citenamefont {Bosak}, \citenamefont {Dai}, \citenamefont {Tallarida}, \citenamefont {Frano}, \citenamefont {Pardo}, \citenamefont {Wu},\ and\ \citenamefont {Blanco-Canosa}}]{Kar2025-jl}%
  \BibitemOpen
  \bibfield  {author} {\bibinfo {author} {\bibfnamefont {A.}~\bibnamefont {Kar}}, \bibinfo {author} {\bibfnamefont {R.}~\bibnamefont {Basak}}, \bibinfo {author} {\bibfnamefont {X.}~\bibnamefont {Li}}, \bibinfo {author} {\bibfnamefont {A.}~\bibnamefont {Korshunov}}, \bibinfo {author} {\bibfnamefont {D.}~\bibnamefont {Subires}}, \bibinfo {author} {\bibfnamefont {J.}~\bibnamefont {Phillips}}, \bibinfo {author} {\bibfnamefont {C.-Y.}\ \bibnamefont {Lim}}, \bibinfo {author} {\bibfnamefont {F.}~\bibnamefont {Zhou}}, \bibinfo {author} {\bibfnamefont {L.}~\bibnamefont {Song}}, \bibinfo {author} {\bibfnamefont {W.}~\bibnamefont {Wang}}, \bibinfo {author} {\bibfnamefont {Y.-C.}\ \bibnamefont {Lau}}, \bibinfo {author} {\bibfnamefont {G.}~\bibnamefont {Garbarino}}, \bibinfo {author} {\bibfnamefont {P.}~\bibnamefont {Gargiani}}, \bibinfo {author} {\bibfnamefont {Y.}~\bibnamefont {Zhao}}, \bibinfo {author} {\bibfnamefont {C.}~\bibnamefont {Plueckthun}}, \bibinfo {author} {\bibfnamefont {S.}~\bibnamefont {Francoual}},
  \bibinfo {author} {\bibfnamefont {A.}~\bibnamefont {Jana}}, \bibinfo {author} {\bibfnamefont {I.}~\bibnamefont {Vobornik}}, \bibinfo {author} {\bibfnamefont {T.}~\bibnamefont {Valla}}, \bibinfo {author} {\bibfnamefont {A.}~\bibnamefont {Rajapitamahuni}}, \bibinfo {author} {\bibfnamefont {J.~G.}\ \bibnamefont {Analytis}}, \bibinfo {author} {\bibfnamefont {R.~J.}\ \bibnamefont {Birgeneau}}, \bibinfo {author} {\bibfnamefont {E.}~\bibnamefont {Vescovo}}, \bibinfo {author} {\bibfnamefont {A.}~\bibnamefont {Bosak}}, \bibinfo {author} {\bibfnamefont {J.}~\bibnamefont {Dai}}, \bibinfo {author} {\bibfnamefont {M.}~\bibnamefont {Tallarida}}, \bibinfo {author} {\bibfnamefont {A.}~\bibnamefont {Frano}}, \bibinfo {author} {\bibfnamefont {V.}~\bibnamefont {Pardo}}, \bibinfo {author} {\bibfnamefont {S.}~\bibnamefont {Wu}},\ and\ \bibinfo {author} {\bibfnamefont {S.}~\bibnamefont {Blanco-Canosa}},\ }\bibfield  {title} {\bibinfo {title} {Magnetoelastic coupling in intercalated transition metal dichalcogenides},\ }\href@noop
  {} {\bibfield  {journal} {\bibinfo  {journal} {arXiv [cond-mat.str-el]}\ } (\bibinfo {year} {2025})}\BibitemShut {NoStop}%
\bibitem [{\citenamefont {Wang}\ \emph {et~al.}(2013)\citenamefont {Wang}, \citenamefont {Li}, \citenamefont {Xiang},\ and\ \citenamefont {Wang}}]{PhysRevB.87.115135}%
  \BibitemOpen
  \bibfield  {author} {\bibinfo {author} {\bibfnamefont {W.-S.}\ \bibnamefont {Wang}}, \bibinfo {author} {\bibfnamefont {Z.-Z.}\ \bibnamefont {Li}}, \bibinfo {author} {\bibfnamefont {Y.-Y.}\ \bibnamefont {Xiang}},\ and\ \bibinfo {author} {\bibfnamefont {Q.-H.}\ \bibnamefont {Wang}},\ }\bibfield  {title} {\bibinfo {title} {Competing electronic orders on kagome lattices at van hove filling},\ }\href@noop {} {\bibfield  {journal} {\bibinfo  {journal} {Phys. Rev. B}\ }\textbf {\bibinfo {volume} {87}},\ \bibinfo {pages} {115135} (\bibinfo {year} {2013})}\BibitemShut {NoStop}%
\bibitem [{\citenamefont {Sherkunov}\ and\ \citenamefont {Betouras}(2018)}]{PhysRevB.98.205151}%
  \BibitemOpen
  \bibfield  {author} {\bibinfo {author} {\bibfnamefont {Y.}~\bibnamefont {Sherkunov}}\ and\ \bibinfo {author} {\bibfnamefont {J.~J.}\ \bibnamefont {Betouras}},\ }\bibfield  {title} {\bibinfo {title} {Electronic phases in twisted bilayer graphene at magic angles as a result of van hove singularities and interactions},\ }\href@noop {} {\bibfield  {journal} {\bibinfo  {journal} {Phys. Rev. B}\ }\textbf {\bibinfo {volume} {98}},\ \bibinfo {pages} {205151} (\bibinfo {year} {2018})}\BibitemShut {NoStop}%
\bibitem [{\citenamefont {Han}\ \emph {et~al.}(2023)\citenamefont {Han}, \citenamefont {Schnyder},\ and\ \citenamefont {Wu}}]{PhysRevB.107.184504}%
  \BibitemOpen
  \bibfield  {author} {\bibinfo {author} {\bibfnamefont {X.}~\bibnamefont {Han}}, \bibinfo {author} {\bibfnamefont {A.~P.}\ \bibnamefont {Schnyder}},\ and\ \bibinfo {author} {\bibfnamefont {X.}~\bibnamefont {Wu}},\ }\bibfield  {title} {\bibinfo {title} {Enhanced nematicity emerging from higher-order van hove singularities},\ }\href@noop {} {\bibfield  {journal} {\bibinfo  {journal} {Phys. Rev. B}\ }\textbf {\bibinfo {volume} {107}},\ \bibinfo {pages} {184504} (\bibinfo {year} {2023})}\BibitemShut {NoStop}%
\bibitem [{\citenamefont {Jiang}\ \emph {et~al.}(2024)\citenamefont {Jiang}, \citenamefont {Shao}, \citenamefont {Xia}, \citenamefont {Denner}, \citenamefont {Ingham}, \citenamefont {Hossain}, \citenamefont {Qiu}, \citenamefont {Zheng}, \citenamefont {Chen}, \citenamefont {Cheng}, \citenamefont {Yang}, \citenamefont {Kim}, \citenamefont {Yin}, \citenamefont {Zhang}, \citenamefont {Litskevich}, \citenamefont {Zhang}, \citenamefont {Cochran}, \citenamefont {Peng}, \citenamefont {Chang}, \citenamefont {Guo}, \citenamefont {Thomale}, \citenamefont {Neupert},\ and\ \citenamefont {Hasan}}]{Jiang2024-te}%
  \BibitemOpen
  \bibfield  {author} {\bibinfo {author} {\bibfnamefont {Y.-X.}\ \bibnamefont {Jiang}}, \bibinfo {author} {\bibfnamefont {S.}~\bibnamefont {Shao}}, \bibinfo {author} {\bibfnamefont {W.}~\bibnamefont {Xia}}, \bibinfo {author} {\bibfnamefont {M.~M.}\ \bibnamefont {Denner}}, \bibinfo {author} {\bibfnamefont {J.}~\bibnamefont {Ingham}}, \bibinfo {author} {\bibfnamefont {M.~S.}\ \bibnamefont {Hossain}}, \bibinfo {author} {\bibfnamefont {Q.}~\bibnamefont {Qiu}}, \bibinfo {author} {\bibfnamefont {X.}~\bibnamefont {Zheng}}, \bibinfo {author} {\bibfnamefont {H.}~\bibnamefont {Chen}}, \bibinfo {author} {\bibfnamefont {Z.-J.}\ \bibnamefont {Cheng}}, \bibinfo {author} {\bibfnamefont {X.~P.}\ \bibnamefont {Yang}}, \bibinfo {author} {\bibfnamefont {B.}~\bibnamefont {Kim}}, \bibinfo {author} {\bibfnamefont {J.-X.}\ \bibnamefont {Yin}}, \bibinfo {author} {\bibfnamefont {S.}~\bibnamefont {Zhang}}, \bibinfo {author} {\bibfnamefont {M.}~\bibnamefont {Litskevich}}, \bibinfo {author} {\bibfnamefont {Q.}~\bibnamefont {Zhang}},
  \bibinfo {author} {\bibfnamefont {T.~A.}\ \bibnamefont {Cochran}}, \bibinfo {author} {\bibfnamefont {Y.}~\bibnamefont {Peng}}, \bibinfo {author} {\bibfnamefont {G.}~\bibnamefont {Chang}}, \bibinfo {author} {\bibfnamefont {Y.}~\bibnamefont {Guo}}, \bibinfo {author} {\bibfnamefont {R.}~\bibnamefont {Thomale}}, \bibinfo {author} {\bibfnamefont {T.}~\bibnamefont {Neupert}},\ and\ \bibinfo {author} {\bibfnamefont {M.~Z.}\ \bibnamefont {Hasan}},\ }\bibfield  {title} {\bibinfo {title} {Van hove annihilation and nematic instability on a kagome lattice},\ }\href@noop {} {\bibfield  {journal} {\bibinfo  {journal} {Nat. Mater.}\ }\textbf {\bibinfo {volume} {23}},\ \bibinfo {pages} {1214} (\bibinfo {year} {2024})}\BibitemShut {NoStop}%
\bibitem [{\citenamefont {Nag}\ \emph {et~al.}(2024)\citenamefont {Nag}, \citenamefont {Batabyal}, \citenamefont {Ingham}, \citenamefont {Morali}, \citenamefont {Tan}, \citenamefont {Koo}, \citenamefont {Consiglio}, \citenamefont {Liu}, \citenamefont {Avraham}, \citenamefont {Queiroz}, \citenamefont {Thomale}, \citenamefont {Yan}, \citenamefont {Felser},\ and\ \citenamefont {Beidenkopf}}]{Nag2024-on}%
  \BibitemOpen
  \bibfield  {author} {\bibinfo {author} {\bibfnamefont {P.~K.}\ \bibnamefont {Nag}}, \bibinfo {author} {\bibfnamefont {R.}~\bibnamefont {Batabyal}}, \bibinfo {author} {\bibfnamefont {J.}~\bibnamefont {Ingham}}, \bibinfo {author} {\bibfnamefont {N.}~\bibnamefont {Morali}}, \bibinfo {author} {\bibfnamefont {H.}~\bibnamefont {Tan}}, \bibinfo {author} {\bibfnamefont {J.}~\bibnamefont {Koo}}, \bibinfo {author} {\bibfnamefont {A.}~\bibnamefont {Consiglio}}, \bibinfo {author} {\bibfnamefont {E.}~\bibnamefont {Liu}}, \bibinfo {author} {\bibfnamefont {N.}~\bibnamefont {Avraham}}, \bibinfo {author} {\bibfnamefont {R.}~\bibnamefont {Queiroz}}, \bibinfo {author} {\bibfnamefont {R.}~\bibnamefont {Thomale}}, \bibinfo {author} {\bibfnamefont {B.}~\bibnamefont {Yan}}, \bibinfo {author} {\bibfnamefont {C.}~\bibnamefont {Felser}},\ and\ \bibinfo {author} {\bibfnamefont {H.}~\bibnamefont {Beidenkopf}},\ }\bibfield  {title} {\bibinfo {title} {Pomeranchuk instability induced by an emergent higher-order van hove singularity on
  the distorted kagome surface of {Co}$_3${Sn}$_2${S}$_2$},\ }\href@noop {} {\bibfield  {journal} {\bibinfo  {journal} {arXiv [cond-mat.str-el]}\ } (\bibinfo {year} {2024})}\BibitemShut {NoStop}%
\bibitem [{\citenamefont {Fradkin}\ \emph {et~al.}(2015)\citenamefont {Fradkin}, \citenamefont {Kivelson},\ and\ \citenamefont {Tranquada}}]{RevModPhys.87.457}%
  \BibitemOpen
  \bibfield  {author} {\bibinfo {author} {\bibfnamefont {E.}~\bibnamefont {Fradkin}}, \bibinfo {author} {\bibfnamefont {S.~A.}\ \bibnamefont {Kivelson}},\ and\ \bibinfo {author} {\bibfnamefont {J.~M.}\ \bibnamefont {Tranquada}},\ }\bibfield  {title} {\bibinfo {title} {Colloquium: Theory of intertwined orders in high temperature superconductors},\ }\href@noop {} {\bibfield  {journal} {\bibinfo  {journal} {Rev. Mod. Phys.}\ }\textbf {\bibinfo {volume} {87}},\ \bibinfo {pages} {457} (\bibinfo {year} {2015})}\BibitemShut {NoStop}%
\bibitem [{\citenamefont {Fernandes}\ \emph {et~al.}(2019)\citenamefont {Fernandes}, \citenamefont {Orth},\ and\ \citenamefont {Schmalian}}]{Fernandes2019-kt}%
  \BibitemOpen
  \bibfield  {author} {\bibinfo {author} {\bibfnamefont {R.~M.}\ \bibnamefont {Fernandes}}, \bibinfo {author} {\bibfnamefont {P.~P.}\ \bibnamefont {Orth}},\ and\ \bibinfo {author} {\bibfnamefont {J.}~\bibnamefont {Schmalian}},\ }\bibfield  {title} {\bibinfo {title} {Intertwined vestigial order in quantum materials: Nematicity and beyond},\ }\href@noop {} {\bibfield  {journal} {\bibinfo  {journal} {Annu. Rev. Condens. Matter Phys.}\ }\textbf {\bibinfo {volume} {10}},\ \bibinfo {pages} {133} (\bibinfo {year} {2019})}\BibitemShut {NoStop}%
\bibitem [{\citenamefont {Teng}\ \emph {et~al.}(2023)\citenamefont {Teng}, \citenamefont {Oh}, \citenamefont {Tan}, \citenamefont {Chen}, \citenamefont {Huang}, \citenamefont {Gao}, \citenamefont {Yin}, \citenamefont {Chu}, \citenamefont {Hashimoto}, \citenamefont {Lu}, \citenamefont {Jozwiak}, \citenamefont {Bostwick}, \citenamefont {Rotenberg}, \citenamefont {Granroth}, \citenamefont {Yan}, \citenamefont {Birgeneau}, \citenamefont {Dai},\ and\ \citenamefont {Yi}}]{Teng2023-ay}%
  \BibitemOpen
  \bibfield  {author} {\bibinfo {author} {\bibfnamefont {X.}~\bibnamefont {Teng}}, \bibinfo {author} {\bibfnamefont {J.~S.}\ \bibnamefont {Oh}}, \bibinfo {author} {\bibfnamefont {H.}~\bibnamefont {Tan}}, \bibinfo {author} {\bibfnamefont {L.}~\bibnamefont {Chen}}, \bibinfo {author} {\bibfnamefont {J.}~\bibnamefont {Huang}}, \bibinfo {author} {\bibfnamefont {B.}~\bibnamefont {Gao}}, \bibinfo {author} {\bibfnamefont {J.-X.}\ \bibnamefont {Yin}}, \bibinfo {author} {\bibfnamefont {J.-H.}\ \bibnamefont {Chu}}, \bibinfo {author} {\bibfnamefont {M.}~\bibnamefont {Hashimoto}}, \bibinfo {author} {\bibfnamefont {D.}~\bibnamefont {Lu}}, \bibinfo {author} {\bibfnamefont {C.}~\bibnamefont {Jozwiak}}, \bibinfo {author} {\bibfnamefont {A.}~\bibnamefont {Bostwick}}, \bibinfo {author} {\bibfnamefont {E.}~\bibnamefont {Rotenberg}}, \bibinfo {author} {\bibfnamefont {G.~E.}\ \bibnamefont {Granroth}}, \bibinfo {author} {\bibfnamefont {B.}~\bibnamefont {Yan}}, \bibinfo {author} {\bibfnamefont {R.~J.}\ \bibnamefont {Birgeneau}},
  \bibinfo {author} {\bibfnamefont {P.}~\bibnamefont {Dai}},\ and\ \bibinfo {author} {\bibfnamefont {M.}~\bibnamefont {Yi}},\ }\bibfield  {title} {\bibinfo {title} {Magnetism and charge density wave order in kagome {FeGe}},\ }\href@noop {} {\bibfield  {journal} {\bibinfo  {journal} {Nat. Phys.}\ }\textbf {\bibinfo {volume} {19}},\ \bibinfo {pages} {814} (\bibinfo {year} {2023})}\BibitemShut {NoStop}%
\bibitem [{\citenamefont {Wu}\ \emph {et~al.}(2024)\citenamefont {Wu}, \citenamefont {Klemm}, \citenamefont {Shah}, \citenamefont {Ritz}, \citenamefont {Duan}, \citenamefont {Teng}, \citenamefont {Gao}, \citenamefont {Ye}, \citenamefont {Matsuda}, \citenamefont {Li}, \citenamefont {Xu}, \citenamefont {Yi}, \citenamefont {Birol}, \citenamefont {Dai},\ and\ \citenamefont {Blumberg}}]{PhysRevX.14.011043}%
  \BibitemOpen
  \bibfield  {author} {\bibinfo {author} {\bibfnamefont {S.}~\bibnamefont {Wu}}, \bibinfo {author} {\bibfnamefont {M.~L.}\ \bibnamefont {Klemm}}, \bibinfo {author} {\bibfnamefont {J.}~\bibnamefont {Shah}}, \bibinfo {author} {\bibfnamefont {E.~T.}\ \bibnamefont {Ritz}}, \bibinfo {author} {\bibfnamefont {C.}~\bibnamefont {Duan}}, \bibinfo {author} {\bibfnamefont {X.}~\bibnamefont {Teng}}, \bibinfo {author} {\bibfnamefont {B.}~\bibnamefont {Gao}}, \bibinfo {author} {\bibfnamefont {F.}~\bibnamefont {Ye}}, \bibinfo {author} {\bibfnamefont {M.}~\bibnamefont {Matsuda}}, \bibinfo {author} {\bibfnamefont {F.}~\bibnamefont {Li}}, \bibinfo {author} {\bibfnamefont {X.}~\bibnamefont {Xu}}, \bibinfo {author} {\bibfnamefont {M.}~\bibnamefont {Yi}}, \bibinfo {author} {\bibfnamefont {T.}~\bibnamefont {Birol}}, \bibinfo {author} {\bibfnamefont {P.}~\bibnamefont {Dai}},\ and\ \bibinfo {author} {\bibfnamefont {G.}~\bibnamefont {Blumberg}},\ }\bibfield  {title} {\bibinfo {title} {Symmetry breaking and ascending in the magnetic
  kagome metal \ce{FeGe}},\ }\href@noop {} {\bibfield  {journal} {\bibinfo  {journal} {Phys. Rev. X}\ }\textbf {\bibinfo {volume} {14}},\ \bibinfo {pages} {011043} (\bibinfo {year} {2024})}\BibitemShut {NoStop}%
\bibitem [{\citenamefont {Farhang}\ \emph {et~al.}(2025)\citenamefont {Farhang}, \citenamefont {Meier}, \citenamefont {Lu}, \citenamefont {Li}, \citenamefont {Wu}, \citenamefont {Mozaffari}, \citenamefont {Madhogaria}, \citenamefont {Zhang}, \citenamefont {Mandrus},\ and\ \citenamefont {Xia}}]{Farhang2025-ob}%
  \BibitemOpen
  \bibfield  {author} {\bibinfo {author} {\bibfnamefont {C.}~\bibnamefont {Farhang}}, \bibinfo {author} {\bibfnamefont {W.~R.}\ \bibnamefont {Meier}}, \bibinfo {author} {\bibfnamefont {W.}~\bibnamefont {Lu}}, \bibinfo {author} {\bibfnamefont {J.}~\bibnamefont {Li}}, \bibinfo {author} {\bibfnamefont {Y.}~\bibnamefont {Wu}}, \bibinfo {author} {\bibfnamefont {S.}~\bibnamefont {Mozaffari}}, \bibinfo {author} {\bibfnamefont {R.~P.}\ \bibnamefont {Madhogaria}}, \bibinfo {author} {\bibfnamefont {Y.}~\bibnamefont {Zhang}}, \bibinfo {author} {\bibfnamefont {D.}~\bibnamefont {Mandrus}},\ and\ \bibinfo {author} {\bibfnamefont {J.}~\bibnamefont {Xia}},\ }\bibfield  {title} {\bibinfo {title} {Discovery of an intermediate nematic state in a bilayer kagome metal \ce{ScV6Sn6}},\ }\href@noop {} {\bibfield  {journal} {\bibinfo  {journal} {arXiv [cond-mat.str-el]}\ } (\bibinfo {year} {2025})}\BibitemShut {NoStop}%
\bibitem [{\citenamefont {Ideue}\ and\ \citenamefont {Iwasa}(2021)}]{Ideue2021-of}%
  \BibitemOpen
  \bibfield  {author} {\bibinfo {author} {\bibfnamefont {T.}~\bibnamefont {Ideue}}\ and\ \bibinfo {author} {\bibfnamefont {Y.}~\bibnamefont {Iwasa}},\ }\bibfield  {title} {\bibinfo {title} {Symmetry breaking and nonlinear electric transport in van der waals nanostructures},\ }\href@noop {} {\bibfield  {journal} {\bibinfo  {journal} {Annu. Rev. Condens. Matter Phys.}\ } (\bibinfo {year} {2021})}\BibitemShut {NoStop}%
\bibitem [{\citenamefont {Yang}\ \emph {et~al.}(2021)\citenamefont {Yang}, \citenamefont {Naaman}, \citenamefont {Paltiel},\ and\ \citenamefont {Parkin}}]{Yang2021-lu}%
  \BibitemOpen
  \bibfield  {author} {\bibinfo {author} {\bibfnamefont {S.-H.}\ \bibnamefont {Yang}}, \bibinfo {author} {\bibfnamefont {R.}~\bibnamefont {Naaman}}, \bibinfo {author} {\bibfnamefont {Y.}~\bibnamefont {Paltiel}},\ and\ \bibinfo {author} {\bibfnamefont {S.~S.~P.}\ \bibnamefont {Parkin}},\ }\bibfield  {title} {\bibinfo {title} {Chiral spintronics},\ }\href@noop {} {\bibfield  {journal} {\bibinfo  {journal} {Nat. Rev. Phys.}\ }\textbf {\bibinfo {volume} {3}},\ \bibinfo {pages} {328} (\bibinfo {year} {2021})}\BibitemShut {NoStop}%
\bibitem [{\citenamefont {Montgomery}(1971)}]{montgomery1971method}%
  \BibitemOpen
  \bibfield  {author} {\bibinfo {author} {\bibfnamefont {H.}~\bibnamefont {Montgomery}},\ }\bibfield  {title} {\bibinfo {title} {Method for measuring electrical resistivity of anisotropic materials},\ }\href@noop {} {\bibfield  {journal} {\bibinfo  {journal} {J. Appl. Phys.}\ }\textbf {\bibinfo {volume} {42}},\ \bibinfo {pages} {2971} (\bibinfo {year} {1971})}\BibitemShut {NoStop}%
\bibitem [{\citenamefont {Dos~Santos}\ \emph {et~al.}(2011)\citenamefont {Dos~Santos}, \citenamefont {De~Campos}, \citenamefont {Da~Luz}, \citenamefont {White}, \citenamefont {Neumeier}, \citenamefont {De~Lima},\ and\ \citenamefont {Shigue}}]{dos2011procedure}%
  \BibitemOpen
  \bibfield  {author} {\bibinfo {author} {\bibfnamefont {C.}~\bibnamefont {Dos~Santos}}, \bibinfo {author} {\bibfnamefont {A.}~\bibnamefont {De~Campos}}, \bibinfo {author} {\bibfnamefont {M.}~\bibnamefont {Da~Luz}}, \bibinfo {author} {\bibfnamefont {B.}~\bibnamefont {White}}, \bibinfo {author} {\bibfnamefont {J.}~\bibnamefont {Neumeier}}, \bibinfo {author} {\bibfnamefont {B.}~\bibnamefont {De~Lima}},\ and\ \bibinfo {author} {\bibfnamefont {C.}~\bibnamefont {Shigue}},\ }\bibfield  {title} {\bibinfo {title} {Procedure for measuring electrical resistivity of anisotropic materials: A revision of the montgomery method},\ }\href@noop {} {\bibfield  {journal} {\bibinfo  {journal} {J. Appl. Phys.}\ }\textbf {\bibinfo {volume} {110}} (\bibinfo {year} {2011})}\BibitemShut {NoStop}%
\bibitem [{\citenamefont {Liu}\ \emph {et~al.}(2024)\citenamefont {Liu}, \citenamefont {Shi}, \citenamefont {Jiang}, \citenamefont {Rosenberg}, \citenamefont {DeStefano}, \citenamefont {Liu}, \citenamefont {Hu}, \citenamefont {Zhao}, \citenamefont {Wang}, \citenamefont {Yao} \emph {et~al.}}]{liu2024absence}%
  \BibitemOpen
  \bibfield  {author} {\bibinfo {author} {\bibfnamefont {Z.}~\bibnamefont {Liu}}, \bibinfo {author} {\bibfnamefont {Y.}~\bibnamefont {Shi}}, \bibinfo {author} {\bibfnamefont {Q.}~\bibnamefont {Jiang}}, \bibinfo {author} {\bibfnamefont {E.~W.}\ \bibnamefont {Rosenberg}}, \bibinfo {author} {\bibfnamefont {J.~M.}\ \bibnamefont {DeStefano}}, \bibinfo {author} {\bibfnamefont {J.}~\bibnamefont {Liu}}, \bibinfo {author} {\bibfnamefont {C.}~\bibnamefont {Hu}}, \bibinfo {author} {\bibfnamefont {Y.}~\bibnamefont {Zhao}}, \bibinfo {author} {\bibfnamefont {Z.}~\bibnamefont {Wang}}, \bibinfo {author} {\bibfnamefont {Y.}~\bibnamefont {Yao}}, \emph {et~al.},\ }\bibfield  {title} {\bibinfo {title} {Absence of ${E}_{2g}$ nematic instability and dominant ${A}_{1g}$ response in the kagome metal \ce{CsV3Sb5}},\ }\href@noop {} {\bibfield  {journal} {\bibinfo  {journal} {Phys. Rev. X}\ }\textbf {\bibinfo {volume} {14}},\ \bibinfo {pages} {031015} (\bibinfo {year} {2024})}\BibitemShut {NoStop}%
\end{thebibliography}%

\end{document}